\documentclass{ws-mpla}
\usepackage[super,compress]{cite}
\usepackage{graphicx}
\usepackage[breaklinks]{hyperref}  
\hypersetup{colorlinks,urlcolor=black,citecolor=black,linkcolor=black,filecolor=black}
\usepackage{breakurl}

 \newcommand{\exclude}[1]{}

\newcommand{\be}{\begin{eqnarray}}
\newcommand{\ee}{\end{eqnarray}}
\def\ra{\rangle}
\def\la{\langle}

\begin{document}

\thispagestyle{plain}

\def\bib{B\kern-.05em{I}\kern-.025em{B}\kern-.08em}
\def\btex{B\kern-.05em{I}\kern-.025em{B}\kern-.08em\TeX}

\title{{\btex}ing in MPLA {\bib}Style}

\exclude{ 
\markboth{WSPC}{Using World Scientific's \btex\/ Style File}
 }

\markboth{Ariel Zhitnitsky}
{Axion Quark Nuggets, Dark Matter and  Matter-Antimatter asymmetry. }

\catchline{}{}{}{}{}

\title{Axion Quark Nuggets. Dark Matter and  Matter-Antimatter asymmetry:  theory, observations and future experiments }

\author{\footnotesize Ariel Zhitnitsky }

\address{University of British Columbia,
Vancouver, BC, V6T1Z1, Canada\\
arz@phas.ubc.ca}

\maketitle

\pub{Received (Day Month Year)}{Revised (Day Month Year)}

\begin{abstract}
We review  a   testable, the axion quark nugget (AQN) model
  outside of the standard WIMP paradigm. The model was originally  invented to explain the  
observed similarity between the dark and the visible components,    $\Omega_{\rm DM}\approx \Omega_{\rm visible}$
  in a   natural way 
as both types of matter  are formed during  the same QCD transition and  proportional to the same  dimensional fundamental parameter of the system, $\Lambda_{\rm QCD}$. In this framework the 
baryogenesis is actually a charge segregation  (rather than charge generation) process  which is operational due to the    $\cal{CP}$-odd axion field,
while   the global baryon number of the Universe remains 
zero.      The   nuggets and anti-nuggets are strongly interacting but macroscopically large objects with approximately nuclear density. We overview  several specific recent applications of this framework. First, we discuss the ``solar corona  mystery" when   the  so-called nanoflares are identified with the AQN annihilation events in corona.  Secondly, we review a proposal that  the recently observed by   the Telescope Array   puzzling events
is a result of the annihilation events of the AQNs under thunderstorm.  Finally, we overview   a  broadband strategy which could lead to the discovery the AQN-induced axions representing  the heart of the construction.

\end{abstract}

\ccode{PACS Nos.: include PACS Nos.}

  \section{Introduction: Motivation}
\label{introduction}
 Two elements  from  the title of this review, 
 the matter-antimatter asymmetry and Dark Matter (DM)   are known to be the two  most challenging problems of the modern cosmology.
Indeed,  we know  about   their existence with great confidence for many years.  However, we still do not know the microscopical nature
of the DM, nor we know why we observe the baryons and not   anti-baryons in the Universe.  

  It is commonly  assumed that there are two separate stories here. The first story which is called the baryogenesis explains  the  matter-antimatter asymmetry in the Universe   as follows. It is believed  that the Universe 
began in a symmetric state with zero global baryonic charge 
and later (through some baryon number violating process, non- equilibrium dynamics, and $\cal{CP}$ violation effects, realizing three  famous   Sakharov's criteria \cite{Sakharov:1967dj}) 
evolved into a state with a net positive baryon number.  The second and independent story  explains the nature of the  DM
in terms of  a new fundamental field which weakly couples to the standard model (SM) particles. For example, it could  be realized    in form of the Weakly Interacting Massive Particles (WIMP)s. 

As an 
alternative to these two separate  stories  we advocate a framework in which 
baryogenesis is actually a charge segregation  (rather than charge generation) process 
when the global baryon number of the universe remains 
zero at all times.  In this model the unobserved antibaryons come to comprise 
the dark matter in the form of dense nuggets of quarks or   antiquarks   in a novel QCD phase, which is an important part of the SM physics.  

The idea that the dark matter may take the form of composite objects of 
standard model quarks in a novel phase goes back to quark nuggets  \cite{Witten:1984rs}, strangelets \cite{Farhi:1984qu}, nuclearities \cite{DeRujula:1984axn}. 
In the models \cite{Witten:1984rs,Farhi:1984qu,DeRujula:1984axn}  the presence of strange quark stabilizes the quark matter at sufficiently 
high densities allowing strangelets being formed in the early universe to remain stable 
over cosmological timescales. 

The axion quark nuggets (AQN)  model, which is the third element from the title of this review, was   advocated in \cite{Zhitnitsky:2002qa} is conceptually similar, with the 
nuggets being composed of a high density colour superconducting (CS) phase.   
As with other high mass dark matter candidates  
  \cite{Witten:1984rs,Farhi:1984qu,DeRujula:1984axn} these objects are ``cosmologically dark" not through the weakness of their 
interactions but due to their small cross-section to mass ratio. 
As a result, the corresponding  constraints on this type of dark matter place a lower bound on their mass, rather than coupling constant.  

There are several additional elements in AQN model in comparison with the older \cite{Witten:1984rs,Farhi:1984qu,DeRujula:1984axn} well-known and well-studied constructions:
\begin{itemize} 
\item
First, there is an additional stabilization factor  provided by the axion domain walls (with a QCD substructure)
which are copiously produced during the QCD transition and which help to alleviate a number of 
the problems inherent in the older versions of the models\footnote{\label{first-order}In particular, in the original proposal the first order phase transition was the required feature of the construction.  However it is known  that the QCD transition is a crossover rather than the first order phase transition. It should be contrasted with  AQN framework
when  the first order phase transition is not required  as the axion domain wall plays the role of the squeezer.  Furthermore, it had been argued that the nuggets   \cite{Witten:1984rs,Farhi:1984qu,DeRujula:1984axn} 
are likely to evaporate on the Hubble time-scale even if they were formed. In the AQN framework 
the  fast evaporation arguments    do not apply    because the  vacuum ground state energies in the  CS  and    hadronic phases   are drastically different. This is because the core of the AQN is in CS phase, which implies that the two systems (CS and hadronic) can coexist only in the presence of the  external pressure which is provided by the axion domain wall. It should be contrasted with original models  \cite{Witten:1984rs,Farhi:1984qu,DeRujula:1984axn} which  must  be stable at zero external pressure. }. 
\item Another  
crucial    additional  element of the AQN proposal      is that the nuggets could be 
made of {\it matter} as well as  {\it antimatter}  during the QCD epoch.  
\end{itemize} 
The direct consequence of the last  feature is that the dark matter  density  
   $\Omega_{\rm DM}$ and the baryonic matter density $ \Omega_{\rm visible}$ will automatically assume the same order of magnitude    $\Omega_{\rm DM}\sim \Omega_{\rm visible}$ without any fine tunings, and irrespectively to  any specific details of the model, such as the axion mass $m_a$  as they are both proportional to the same fundamental $\Lambda_{\rm QCD} $ scale,  
and they both are originated at the same  QCD transition.
  If these processes 
are not fundamentally related the two components 
$\Omega_{\rm DM}$ and $\Omega_{\rm visible}$  could easily 
exist at vastly different scales. 
 Precisely this fundamental consequence of the model was the main  motivation for its original construction.
 The main characteristic  of a nugget is its    baryon charge  $ B \sim R^3$ or what is the same,  its  mass  $M\sim B$ as both parameters are expressed  in terms of the nugget's  size $R$. All observables will be formulated  in terms of the AQN's baryon charge $B$. 
  
 Unlike conventional dark matter candidates, such as WIMPs 
  the AQNs
  are strongly interacting and macroscopically large.  
However, they do not contradict any of the many known observational
constraints on dark matter or
antimatter  due to the following  main reasons~\cite{Zhitnitsky:2006vt}:
\begin{itemize} 
\item They are absolutely stable configurations on cosmological scale as the pressure due to the axion domain  wall (with the QCD substructure)   is equilibrated by the Fermi pressure. Furthermore, it has been shown that the AQNs survive an unfriendly environment of early Universe, before and after BBN epoch. The majority of the AQNs also survive such violent events as  the galaxy formation and star formation \cite{Ge:2019voa};
\item 
  The nuggets in CS phase have approximately the   nuclear density.  Furthermore, their effective cross section $\sigma\sim R^2\sim B^{2/3}$
 determines   the key ratio $\sigma/M\sim B^{-1/3} \ll 1$   entering all the observables.
This ratio in the AQN model is   well below the typical astrophysical and cosmological limits which are on the order of 
$\sigma/M\lesssim 1$~cm$^2$/g. This is precisely the reason why the strongly interacting AQNs are qualified candidates to serve as   the DM particles;
\item They have a large binding energy with a typical for CS phases gap $\Delta\gtrsim 40$ MeV, such that the baryon charge locked in the
nuggets is not available to participate in big bang nucleosynthesis
(BBN) at $T \sim 0.1$~MeV, and the basic BBN picture holds, with possible small deviations  
of order $\sim 10^{-10}$ which may in fact resolve the primordial lithium puzzle \cite{Flambaum:2018ohm};
\item The nuggets completely decouple from photons as a result of small $ \sigma/M \sim B^{-1/3}\sim 10^{-10} ~\rm cm^2/g$ ratio, such that conventional picture for structure formation holds. Due to the same reasons, 
  the nuggets do not modify conventional CMB analysis. 
\end{itemize} 
 To reiterate: the weakness of the visible-dark matter interaction is achieved 
in this model due to the small geometrical factor $ {\sigma}/{M} \sim B^{-1/3}$ 
rather than due to a weak coupling of a new fundamental field to standard model particles. 
In other words, this small effective interaction $\sim \sigma/M \sim B^{-1/3}$ 
replaces a conventional requirement $\sigma/M\ll 1$~cm$^2$/g
of sufficiently weak interactions of the visible matter with WIMPs.

The review is organized as follows. In Section \ref{formation} we  highlight the AQN  formation mechanism  during the QCD epoch.  
We also make few comments on the size distribution and corresponding observational constraints.  Three next  sections which follow    are devoted to the applications of the AQN framework to the observations, predictions, and possible future experiments. To be more specific, the  section
 \ref{corona}  is  devoted to the so-called ``solar corona mystery" and how it could be naturally resolved within  the AQN framework. 
 The section \ref{TA-bursts} is  devoted to explanation of the recently   observed by the Telescope Array (TA) puzzling events, the so-called  bursts.
These events are very unusual and  drastically different from  conventional cosmic ray (CR) air  showers. We explain how these puzzling   features   could be naturally explained within the AQN framework. Finally,  section \ref{axion}   highlights the basic ideas  on possible  broadband strategy to search for the   AQN-induced axions.
 
\section{Formation mechanism}\label{formation}

This section represents a short overview of the AQN formation mechanism. We refer to the original papers ~\cite{Liang:2016tqc,Ge:2017ttc,Ge:2017idw}  for the technical details   by highlighting the basic conceptual ideas below.    
As we mentioned in Introduction the baryogenesis is replaced by ``charge segregation" mechanism in this framework.  
The result of this   process is two populations of AQN carrying positive and 
negative baryon charges. In other words,  the AQN may be formed of either matter or antimatter. 
However, due to the global  $\cal CP$ violating processes associated with initial misalignment angle $\theta_0\neq 0$ during 
the early formation stage,  a typical  baryon charge hidden in  nuggets  $B_{\rm N}$ and antinuggets $B_{\rm \bar{N}}$ 
  will be different\footnote{This source of strong ${\cal CP}$ violation is no longer 
available at the present epoch as a result of the dynamics of the axion, 
 which  remains the most compelling resolution of the strong ${\cal CP}$ problem, see original papers 
 on PQ symmetry \cite{1977PhRvD..16.1791P}, Weinberg-Wilczek axion \cite{1978PhRvL..40..223W,1978PhRvL..40..279W},
 KSVZ invisible  axion \cite{KSVZ1,KSVZ2} and DFSZ invisible axion  \cite{DFSZ1,DFSZ2} models. See also  recent reviews \cite{Marsh:2015xka,Graham:2015ouw,Irastorza:2018dyq,Sikivie:2020zpn}.}.
 This difference  is always an order of one effect  as expressed by parameter $c\sim1 $  in (\ref{Omega}) below. This  effect  occurs irrespectively to the 
parameters of the theory, the axion mass $m_a$ or the initial misalignment angle $\theta_0$.
   The resulting  disparity  between nuggets $\Omega_{N}$ and antinuggets $\Omega_{\bar{N}}$  generated by the   $\cal CP$ violation  unambiguously implies that  the  baryon contribution $\Omega_{B}$ must be the same order of magnitude as $\Omega_{\bar{N}}$  and $\Omega_{N}$ because all these components are proportional to one and the same fundamental dimensional parameter $\Lambda_{\rm QCD}$ as all dimensional parameters in QCD such as the CS gap $\Delta$, critical temperature $T_c$, chemical potential $\mu$ always assume  the same order of magnitude as $\Lambda_{\rm QCD}$, see \cite{Liang:2016tqc} with the details. 
The remaining antibaryons in the early universe plasma then 
annihilate away leaving only the baryons whose antimatter 
counterparts are bound in the excess of antiquark nuggets and are thus 
unavailable for fast annihilation.  As all  asymmetry effects   are order of
one it eventually  results  in similarities for all 
  components, visible and dark, i.e.
\be
\label{Omega}
\Omega_{\rm DM} \approx (\Omega_{\rm N}+\Omega_{\rm \bar{N}}), ~~ \Omega_{\rm DM}\approx\left(\frac{1+c}{1-c}\right)\Omega_{\rm B}, ~~ (B_{\rm N}+B_{\rm \bar{N}}+B_B)=0, ~~      c\equiv\frac{|B_{\rm \bar{N}}|}{|B_{\rm N}|},~~
\ee
as they are both proportional to the same fundamental $\Lambda_{\rm QCD} $ scale,  
and they both are originated at the same  QCD epoch.  This represents a  precise mechanism of how the  ``charge segregation"  processes  in the AQN framework replaces the baryogenesis in conventional paradigm. 
 In particular, the observed 
matter to dark matter ratio $\Omega_{\rm DM} \approx 5\cdot \Omega_{\rm B}$ 
corresponds to a scenario in which the baryon charge hidden in  antinuggets  
  is larger than the  baryon charge hidden in nuggets     by a factor of $ c\approx (\Omega_{\bar{N}}/ \Omega_{N}) \approx$ 3/2 at the end of nugget formation. 

It is important to emphasize that the AQN mechanism is not sensitive to the axion mass $m_a$ and it is capable to saturate observable ratio $\Omega_{\rm DM}  \approx 5\cdot \Omega_{\rm B}$ itself without any other additional contributors\footnote{This is because the formation mechanism of the AQN is entirely based on QCD physics, not the axion physics. The axion field enters the formation stage exclusively   in terms of the $\cal CP$ violating phase by generating the disparity  between nuggets $\Omega_{N}$ and antinuggets $\Omega_{\bar{N}}$, as explained above.}. It should be contrasted with conventional axion production mechanisms when the corresponding  contribution  scales as $\Omega_{\rm axion}\sim m_a^{-7/6}$. This scaling  unambiguously implies that the axion mass must be fine-tuned  $m_a\lesssim 10^{-5} {\rm eV}$
 to  saturate the DM density today  while larger axion mass will contribute very little to $\Omega_{\rm DM}$. The relative role  between the direct axion contribution $\Omega_{\rm axion}$ and the AQN contribution to $\Omega_{\rm DM}$ as a function of mass $m_a$ has been studied in \cite{Ge:2017idw}, see Fig. 5 in that paper.

 Another fundamental ratio (along with 
$\Omega_{\rm DM} \sim  \Omega_{\rm B}$  discussed above)
is the baryon to entropy ratio at present time
\be
\label{eta1}
\eta\equiv\frac{n_B-n_{\bar{B}}}{n_{\gamma}}\simeq \frac{n_B}{n_{\gamma}}\sim 10^{-10}.
\ee
If the nuggets were not present after the QCD transition the conventional baryons 
and anti-baryons would continue to annihilate each other until the temperature 
reaches $T\simeq 22$ MeV when density would be 9 orders of magnitude smaller 
than observed (\ref{eta1}). This annihilation catastrophe, normally thought   to be  resolved   as a result of  baryogenesis as formulated by Sakharov \cite{Sakharov:1967dj}.  
In contrast, in the AQN framework  this ratio 
is determined by the formation temperature $T_{\rm form}\simeq 41 $~MeV  at which the nuggets and 
antinuggets complete their formation, when all visible anti-baryons get annihilated and only the visible baryons remain in the system. 
The $T_{\rm form}$ is very hard to compute theoretically as even the phase diagram for CS phase is not well known.  This temperature  of cosmic plasma   is  known with high precision from  the observed  ratio (\ref{eta1}).  However, we  note that $T_{\rm form}\approx \Lambda_{\rm QCD}$   assumes a typical QCD value, as it should as there are no any small parameters in QCD.

The next conceptual question  we want to mention here   is related to the axion domain wall (DW)   formation during the QCD transition,
which represents a key element of the construction. 
There is a subtle  point here which can be explained as follows. 
It is normally assumed that the topological defects cannot be formed 
if there is a unique vacuum state.  At the same time it is assumed that  the Peccei-Quinn (PQ) phase 
transition in the AQN framework occurs before the inflation.   Normally, in this case no topological 
defects can be formed   as there is a single physical vacuum state which occupies entire   observable Universe. 
However, the  so-called  $N_{\rm DW}=1$  
domain wall solution still exists when the system is characterized by unique vacuum state. The  subtle  point here is that the non-trivial solution 
  interpolates between one and 
the same physical but topologically distinct  vacuum states, i.e. $\theta\rightarrow \theta +2\pi k$, similar to well known solitons in the sine-Gordon model. These different topological sectors being 
  classified by integer parameter  $k$   must be   present in the system in every point of space-time, and inflation cannot  separate different topological sectors as a result of  expansion. Therefore $N_{\rm DW}=1$ can be formed even if the PQ phase transition happened before inflation, see \cite{Liang:2016tqc} with the  technical  details. 

\begin{figure}
    \centering
    \includegraphics[width=4.0in]{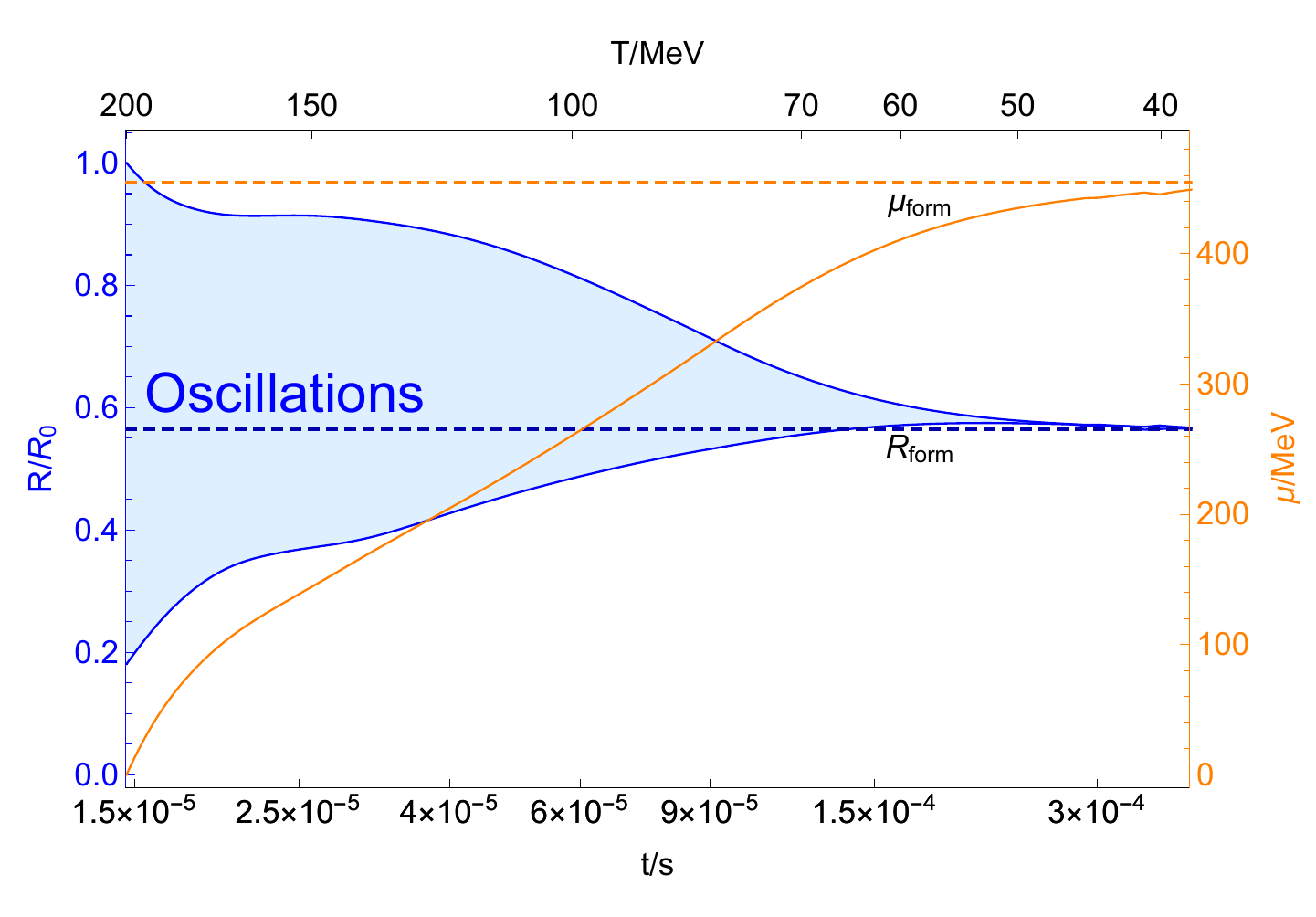}
    \caption{Numerical result of the nugget evolution, adopted from \cite{Ge:2019voa}, see items 1-3 in the text with explanations of the notations. }
    \label{fig:RTE}
\end{figure}

The  numerical simulations suggest \cite{Vilenkin1994} that approximately $87\%$ of the total wall area belong to the percolated large cluster, while the rest $13\%$ is represented by relatively small closed bubbles of different topology. This implies  that a finite  portion  of order $0.1$ of the $N_{\rm DW}=1$ walls are formed as closed surfaces.
  The collapse of the closed $N_{\rm DW}=1$ bubbles will be halted due to the 
Fermi pressure acted by the accumulated fermions. As a result,  the closed 
$N_{\rm DW}=1$ bubbles will eventually become the stable AQNs  and serve as the 
dark matter candidates. The corresponding evolution is rather complex as it includes three drastically different scales: the 
$\Lambda_{\rm QCD}$, the  axion mass scale $m_a$, and finally the cosmological time evolution scale $ t   \sim 10^{-4} $ s when the AQN formation occurs\footnote{The evolution of the Universe with these 3 dramatically different scales must be contrasted with heavy ion experiments when there is a single scale of the problem, the $\Lambda_{\rm QCD}$. Nevertheless, if one translates the achieved in heavy ion collision experiment temperature $T\approx 200$ MeV to cosmological time it would correspond to $ t   \sim 10^{-5} $ s.}.
 The key elements  of this complicated evolution can be summarized as follows:

1. the nugget oscillates  numerous number of times  with frequency $\omega\sim R^{-1}$ by slowly approaching its final  size $R_{\rm form}$,  see shaded blue region 
 on  Fig. \ref{fig:RTE};
 
 2. the nugget  assumes its final configuration with size $R_{\rm form}$  at $T_{\rm form}\approx 40 ~{\rm MeV}$,  see dashed blue line on Fig. \ref{fig:RTE}. This magnitude for $T_{\rm form}$   is consistent with observed value for $\eta$ defined by (\ref{eta1});  
 
3.  the chemical potential inside the nugget   assumes a sufficiently large value  $\mu_{\rm form}\gtrsim 450$~MeV during this long evolution, see orange line on Fig. \ref{fig:RTE}. This magnitude  is consistent with  formation of a 
  CS phase. Therefore, the    original assumption on CS phase which was used in construction of the nugget is justified a posteriori.
 
 We conclude this section with the following comments regarding the formation and survival pattern of the AQNs.  First of all, a  complete formation of the nugget occurs on a time scale $10^{-4} {\rm s}$ which is precisely the cosmological scale when the temperature drops   to $40~ {\rm MeV}$, see Fig. \ref{fig:RTE}. This scale is known from completely different arguments \cite{Kolb:1990vq} related to the estimate of the baryon to photon ratio (\ref{eta1}).   It is a highly nontrivial observation that all these drastically different scales as mentioned above, nevertheless lead to a consistent picture.  Secondly, the newly formed nuggets survive an unfriendly environment of a very hot and dense cosmic plasma before and after BBN epoch at $T\sim 0.1 \rm ~MeV$. Furthermore, a long-standing primordial lithium puzzle may find its resolution within the AQN framework as argued in \cite{Flambaum:2018ohm}. Third, 
the dominant portion of the nuggets survive   the dramatic events (such as galaxy and star formation) during the long post-recombination evolution of the Universe as argued in \cite{Ge:2019voa}. 

The space  limitation   here  does not allow  to cover all these interesting topics in the present review. 
Instead, we quote the known  observational constraints on such kind of objects. 
Any  direct or indirect detection experiments is sensitive to the average value $\langle B \rangle$  
 for the distribution of the AQNs as any observable consequences 
will be scaled by the matter-AQN interaction rate along a given line of sight,
\begin{equation}
\label{eq:observable}
    R^2\int d\Omega dl [n_{\rm visible}(l)\cdot n_{ \rm DM}(l)] \sim \frac{1}{\langle B \rangle^{1/3}},
\end{equation}
where $R\sim B^{1/3}$ and $n_{DM}\sim B^{-1}$ such that effective interaction is suppressed as $  B^{-1/3}$ for large nuggets,    which  represents very generic feature of the model as  discussed   at the very end of  Section \ref{introduction}. 
Thus,  any astrophysical constraints impose a lower bound on the value of 
$\langle B \rangle$. 
The relevant constraints come from a variety of both direct detection and astrophysical observations. 
 
 The strongest direct detection limit 
is  set by the IceCube Observatory's non-detection of a nugget flux which can be expressed as 
\be
\label{direct}
\la B \ra > 3\cdot 10^{24} ~~~~~({\rm direct ~ detection ~constraint)},
\ee
see Appendix A in \cite{Lawson:2019cvy}. 
 Similar limits are also 
derivable from the Antarctic Impulsive Transient Antenna  (ANITA) \cite{Gorham:2012hy} though 
this result   depends  on the details of radio band emissivity of the AQN.   There is also a limit \cite{Gorham:2012hy} from potential contribution to earth's energy budget which  require $|B| > 2.6\times 10^{24}$, which is consistent with (\ref{direct})\footnote{There is also a 
constraint on the flux of heavy dark matter with mass $M<55$g based on the non-detection of 
etching tracks in ancient mica \cite{Jacobs:2014yca}. This constrained is obtained  under assumption  that all nuggets have the same mass, which is not the case for the AQN model as we discuss later in  the next sections.}.

 \section{The AQN model:  application to the  ``solar corona  mystery"}
 \label{corona} 
In this section we review several recent papers  \cite{Zhitnitsky:2017rop,Zhitnitsky:2018mav,Raza:2018gpb, Ge:2020xvf}    devoted to a possible resolution of the long standing problem, the so-called ``solar corona  mystery". We start in subsection \ref{nanoflares} by reviewing the puzzling observations from corona, while in subsection \ref{identification} we explain how these mysterious  features could be naturally explained within the AQN framework. Finally, in subsection \ref{radio} we   interpret    the recently observed   radio impulsive events in quiet solar corona  as inevitable  consequence of  the AQN annihilation events.

\subsection{The nanoflares: the observations and modelling}\label{nanoflares} 
We start by explaining the puzzling features of the solar corona
which  is a very peculiar environment.  Indeed,  at an altitude of 2000 km above of the photosphere, the plasma temperature exceeds a few $10^6$ K. The total energy radiated away by the corona is of the order of $L_{\rm corona} \sim 10^{27} {\rm erg~ s^{-1}}$, which is about $(10^{-6}-10^{-7})$ of the total energy radiated by the Sun. Most of this energy is radiated at the extreme ultraviolet (EUV) and soft X-ray wavelengths.    
There is a very sharp (relatively thin, 200 km at most) transition region (TR)   where the temperature suddenly jumps from $6\cdot 10^3$ K to $10^6$ K. 
This jumps is very uniform and occurs everywhere even in the quiet Sun, where the magnetic field is small,
 ($\sim 1~ {\rm G}$), away from active spots and coronal holes. It is very hard to imagine how the temperature   increases by factor $10^2$ or so 
  over entire surface when the  density  decreases. These dramatic changes occur      on a  relatively short length scale   $\sim 10^2$ km, while a  
    typical   scale in the Sun  of order $\sim 10^5$ km.

A possible solution to the heating problem in the quiet Sun corona was proposed in 1983 by Parker \cite{Parker}, who postulated that a continuous and uniform sequence of miniature flares, which he called ``nanoflares'', could happen in the corona. 
 The term ``nanoflare" has been used in a series of papers by Benz and coauthors
  \cite{Benz-2000,Kraev-2001,Benz-2002, Benz-2003}, and many others\cite{Pauluhn:2006ut,Bingert:2013,Klimchuk:2005nx,Klimchuk:2017}  to advocate the idea that these small ``nano-events'' might be  responsible for the   heating of the   quiet solar corona. Originally, the nanoflares thought to be a ``nano"- version of  large solar flares when the energy is assumed to be generated by the magnetic field reconnection.  
  However, it is very hard to imagine how the magnetic reconnection could occur in the quiet Sun, where the magnetic field  pressure $\sim {\cal{B}}^2$ is four orders of magnitude smaller than conventional kinetic pressure $p$. Therefore, 
  more recently,  the nanoflares are modelled as invisible (below the instrumental threshold) generic events, producing an impulsive energy release at a small scale without specifying  their cause and their nature, see reviews  \cite{Klimchuk:2005nx,Klimchuk:2017}.  
 The list of puzzling features includes:

 1. The EUV emission is highly isotropic \cite{Benz-2002,Benz-2003}, in huge contrast with flares which are much more energetic and occur exclusively in active areas. Therefore the nanoflares have to be distributed very uniformly everywhere, including large areas of  quiet regions; 
  
2. According to \cite{Kraev-2001}, in order to reproduce the measured EUV excess, the observed range of nanoflares  needs to be extrapolated from the observed events  interpolating between $(3.1\cdot 10^{24}  - 1.3\cdot 10^{26})~{\rm erg}$ to unresolved events with energies  $ 10^{22}$ erg and even lower;

 \exclude{
3. The nanoflares and microflares appear in a different range of temperature and emission measure (see  Fig.3 in \cite{Benz-2003}). While  the instrumental  limits prohibit observations at intermediate temperatures, nevertheless the authors of \cite{Benz-2003} argue that  ``the occurrence rates of nanoflares and microflares are so different that they cannot originate from the same population". We emphasize on this difference to argue that the flares originate at sunspot areas with locally large magnetic fields ${\cal{B}}\sim (10^2-10^3)$ G, while  the EUV emission (which is observed  even in very quiet regions where the magnetic  field  is in the range ${\cal{B}}\sim 1$G) is isotropic and covers the entire solar surface;
}
3. Time measurements of many nanoflares demonstrate the Doppler shift with a typical velocities (250-310) km/s,   far exceeds the thermal ion velocity which is around 11 km/s   \cite{Benz-2000};

 4. The EUV emission from the corona shows  a   modest variation  during the solar cycles, not exceeding factor (3-4). It should be contrasted  with  enormous fluctuations  $\sim 10^2$ of the large flare's  frequency of appearance during the same cycles, see Fig. 1 from\cite{Bertolucci-2017}.  

 If the magnetic reconnection  were  fully responsible for both the large flares and nanoflares, then  the variation during the solar cycles should be similar for these two phenomena. It is not what is observed:  the  variation of the EUV  
  during the solar cycles is relatively  modest and  does not normally exceed factor of 3 or so, while 
 the variation of the flare's activity during the same solar cycles very often  exceeds factor $10^2$. 
Therefore,  the source of the uniform  and persistent  EUV radiation must be very different from  the magnetic field activity responsible for the large flares. In particular, the EUV emission  from corona never stops even when the  large flare activity is not observed for months. The source of this persistent  EUV radiation still remains   a big mystery.

  \exclude{
 The nanoflares are usually characterised by the following distribution:
 \begin{equation}
\label{dN}
\rm d N\propto   W^{-\alpha}\rm d W ~~~~ 10^{21}{\rm erg}
 \lesssim W\lesssim 10^{26} {\rm erg}
\end{equation}
 where $\rm d N$ is the number of nanoflare events per unit time, with energy between $W$ and $W+\rm d W$. 
 In formula (\ref{dN}) we display  the approximate energy window for $W$ as expressed by items {\bf 2} and {\bf 3} including the sub-resolution events extrapolated to very low energies. 
 The distribution   $\rm d N/\rm d W$ has been  modelled via magnetic-hydro-dynamics (MHD) simulations \cite{Pauluhn:2006ut,Bingert:2013} in such a way that the Solar observations match simulations. The parameter $\alpha$ was  fixed to fit  observations \cite{Pauluhn:2006ut,Bingert:2013}, see description of different models in next subsection.  
}

 \subsection{The nanoflares as the AQN annihilation events}\label{identification} 
 In this subsection we explain how these observed puzzling features listed above are naturally occur in the AQN framework.  
 It has been conjectured in  \cite{Zhitnitsky:2017rop} that the 
  nanoflares can be identified with the AQN annihilation events when the nuggets hit the sun and release their entire energy to corona.    From this identification it follows that the total AQN's annihilating charge should equal to the  energy of a nanoflare. Furthermore, the baryon charge distribution within AQN framework and the nanoflare energy distribution must be one and the same function \cite{Zhitnitsky:2017rop}, i.e.
  \begin{equation}
\label{W_B}
  d N\propto B^{-\alpha}  d B\propto W^{-\alpha}  d W  \iff  [\rm nanoflares\equiv AQN ~ annihilation~ events], 
\end{equation}
where $  d N$ is the number of nanoflare events  with energy between $W$ and $W+  d W$, which occur as a result of complete annihilation of the antimatter AQN carrying the baryon charges between $B$ and $B+  d B$.  

An immediate
self-consistency check of this conjecture is the observation that the allowed   constraint  (\ref{direct})  for the AQNs baryonic charge $B$ is consistent  with  the     nanoflare energy  $W$   as these two values become connected  as  $W\approx 2~ {\rm GeV}\cdot B$. In particular,  the minimal baryon charge $B_{\rm min}\simeq 3\cdot 10^{24}$ from (\ref{direct}) corresponds to the released energy by a nanoflare $W_{\rm min}\simeq (2~ {\rm GeV} ) B_{\rm min}\simeq 10^{22}$ erg, which is a proper scale for the extrapolation 
to the unresolved events as mentioned in item 2 from previous subsection \ref{nanoflares}. 
  One should emphasize that this is a highly nontrivial self-consistency check of the proposal  \cite{Zhitnitsky:2017rop} as the acceptable range   for the AQNs and nanoflares   have been constrained from drastically different  physical  systems. 
  
  We are now in position to present several additional arguments to support this proposal: item 1 from previous subsection \ref{nanoflares}  is also naturally explained in the AQN framework as the DM is expected to be distributed very uniformly over the Sun making no distinction between quiet and active regions, in contrast with large flares. A similar argument applies to item 4   as the strength of the magnetic field and its localization is absolutely irrelevant for the nanoflare events in form of the AQNs, in contrast with conventional paradigm when the nanoflares are thought to be simply   scaled down configurations  of their larger cousin  which are much more energetic and occur exclusively in active areas and cannot be uniformly distributed. 
 It is also consistent with observation that  the temporal modulation of the EUV irradiance over a solar cycle is very small and does not exceed a factor $\sim 3$, as opposed to the much dramatic changes in Solar activity with modulations on the level of $10^2$ over the same time scale. This suggests that the energy injection from the nanoflares is weakly related to the Solar activity.
   \begin{figure}
    \centering
    \includegraphics[width=4.0in]{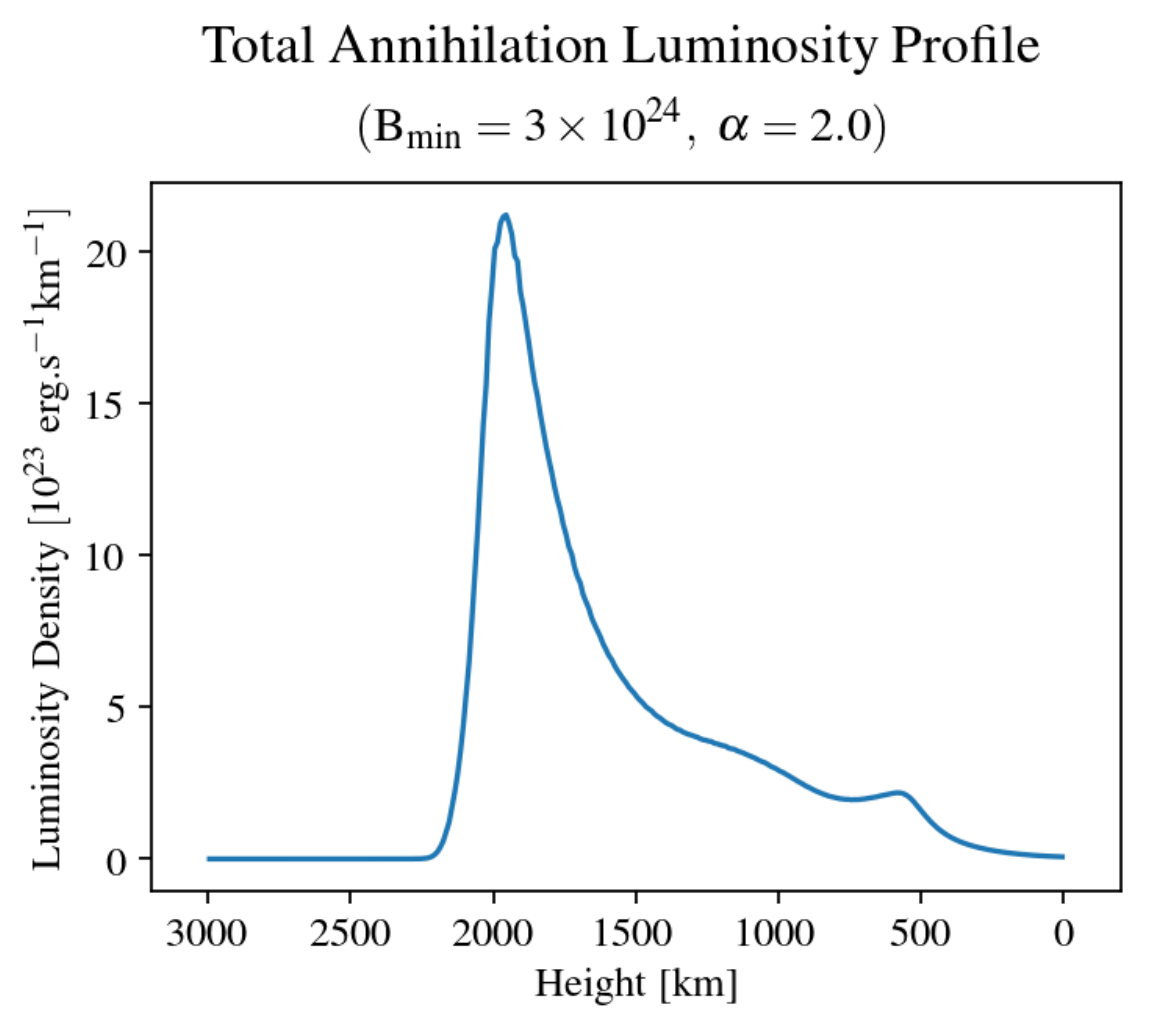}
    \caption{Total EUV luminosity as a function of height, adopted from \cite{Raza:2018gpb}. It is a highly nontrivial consequence of the conjecture (\ref{W_B}) that the most of the energy is released in the narrow TR around 2000 km, where it is known that the most of the drastic changes occur.}
    \label{luminosity}
\end{figure}
 The presence of the large Doppler shift  with a typical velocities (250-310) km/s, mentioned in  item 3, can be  understood  within the AQN framework as follows: the typical velocities of the nuggets entering the solar corona is very high, around 700 km/s.  The Mach number  $M=v_{\rm AQN}/c_s$  is also very large. A shock wave will be inevitably formed and will push the surrounding material to the velocities which are much higher than would normally present in the  thermal equilibrium with its velocity on the level $11 \rm km/s$.

  Encouraged by these self-consistency checks    the authors of  \cite{Raza:2018gpb} literally  used the distribution function $f(W)$ 
  which was previously used   in \cite{Pauluhn:2006ut,Bingert:2013} to fit the solar data.  This step represents a precise realization of the identification
  (\ref{W_B}) 
  with power-law index $\alpha$ which  normally lies in the window $\alpha\in(2-2.5)$ to fit the corona observations.  This identification  allows  
     to describe the AQN's baryon number distribution 
 $dN/dB$. It can be used for the solar corona as well as for   any other applications\footnote{One should   note that it has been  argued \cite{Ge:2019voa} that the algebraic scaling (\ref{f(B)}) is a generic feature of the AQN formation mechanism based on percolation theory. The phenomenological parameter $\alpha$ is determined by the properties of the domain wall formation during the QCD transition in the early Universe, but it cannot be theoretically computed in strongly coupled QCD. Instead, it is constrained by the observations as discussed in the text.}. 
 \begin{equation}
\label{f(B)}
\frac{ dN}{dB}\propto     f(B)\propto B^{-\alpha}, ~~~~ \langle B\rangle 
=\int_{B_{\rm min}}^{B_{\rm max}}  d B~[B ~ f(B)],~~~\int_{B_{\rm min}}^{B_{\rm max}}  d B~ f(B)=1,
\end{equation}
where $B_{\rm min}$ and $B_{\rm max}$ were also fixed using the identification (\ref{W_B}).
 
 With explicit models for the distribution function $f(W)$  and with absolute  normalization for the AQN's  flux determined by the DM density $\rho_{\rm DM}\approx 0.3~ \rm GeV/cm^3$ one can proceed with Monte Carlo (MC) simulations to study a number of interesting questions related to the dynamics of the AQNs when they propagate in solar atmosphere. The results of the corresponding studies performed in \cite{Raza:2018gpb} were remarkable: the total energy rate injection  into the solar corona as a result of the AQN annihilation processes is very close to the observed  value $L_{\rm corona} \sim 10^{27} {\rm erg~ s^{-1}}$. Furthermore, this energy is mostly deposited in the TR around 2000 km as mentioned in Section \ref{nanoflares}, see Fig.\ref{luminosity}.  It  could   explain  the nature of this region   when very sharp changes in temperature occur on a very small scale as a result of the AQN annihilation processes in this transition region\footnote{\label{TR}To the best of my  knowledge most or even all made proposals to solve the corona problem cannot explain the narrow transition region of the solar atmosphere. Here it emerges in a very natural way without fitting of any parameters.}.

\subsection{Impulsive radio events in quiet solar corona}\label{radio}
As discussed in the previous subsections the individual short energy bursts associated with these nanoflares are   below detection limits and have not yet been directly observed in the EUV or  X-ray regimes. In fact, all coronal heating models advocated so far, including \cite{Pauluhn:2006ut,Bingert:2013}    require the existence of an unobserved (i.e. unresolved with current instrumentation) source of energy distributed over the entire Sun.  However, it has been recently claimed  in \cite{Mondal-2020}  that radio observations   can potentially ``see" individual nanoflares and their ``internal structures", where more energetic EUV and X-rays instruments cannot, while  the intensity at the  radio frequencies is many orders of magnitude smaller than more energetic EUV and X-rays radiation.  

In our recent paper \cite{Ge:2020xvf}   it has been argued that the radio emission fits in the AQN annihilation events. 
It has been also argued  that    
the impulsive radio events in the quiet solar corona as recorded by the Murchison Widefield Array (MWA) \cite{Mondal-2020} match 
well the computations based on     the  AQN
    framework.  This claim has been supported by demonstrating that the generic features of the observations in  the  radio frequency bands  such as   the rate of appearance, the temporal and spatial distributions and their energetics  represent the   natural  consequences    of the AQN annihilation events  in the quiet corona.
    
    The basic idea of the radio emission due to the AQN annihilation events can be briefly explained as follows. It is generally accepted that the radio emission from the corona results from the interaction of plasma  oscillations (also known as Langmuir waves) with non-thermal electrons  which must be injected into the plasma by some non-thermal mechanism\cite{Thejappa-1991}.   The  plasma instability develops  when the injected electrons have a  non-thermal high energy component  at which point the  radio waves  can be emitted. 
The frequency of emission $\nu$ is mostly determined by the plasma frequency $\omega_p$ in a given environment, i.e. 
 \begin{equation}
\label{eq:omega}
 \omega^2=\omega_p^2+ k^2\frac{3T}{m_e}, ~~~    \omega_p^2=\frac{4\pi\alpha n_e}{m_e},  ~~~ \nu=\frac{\omega}{2\pi}, 
\end{equation}
where $n_e$ is the electron number density in the corona at altitude $h$,  while $T$ is the temperature at the same altitude and $k$ is the wavenumber.
The emission of radio waves generically occurs at high  altitudes $h\sim 10^4$ km  where the energy is injected into the plasma. 
The non-thermal electrons  received their kinetic energy at much lower  altitude $h\simeq 2000$ km where the AQNs release their annihilation energy as discussed in previous subsection, see Fig.\ref{luminosity}. After that they can propagate upward for  very long distance determined by the mean free path $\lambda\sim 10^4$ km before they transfer their energy to the radio waves.  It is known that the number density of the non-thermal (supra-thermal in terminology  \cite{Thejappa-1991}) electrons $n_{s}$   must be sufficiently large  $n_{s}/n_e\gtrsim  10^{-7}$ for the plasma instability to develop, in which  case   the radio waves will be generated \cite{Thejappa-1991}.  As the density $n_{s}/n_e$ approaches the threshold values at some specific frequencies, the intensity increases sharply, which we identify with the observed impulsive radio events. These threshold conditions may be satisfied randomly in space and time, depending on properties of the injected electrons  and properties of the surrounding plasma\cite{Thejappa-1991}.  

It has been  demonstrated in   \cite{Ge:2020xvf}  that 
  the number density of the non-thermal electrons $n_{s}$  generated by the AQNs can easily  assume a proper range $n_{s}/n_e\gtrsim  10^{-7}$  for the plasma instability to develop when these     non-thermal electrons reach the altitudes $h\sim 10^4$ km where the radio emission occurs as  the plasma frequency $\omega_p(h)$ assumes a proper value (\ref{eq:omega}). It has been also shown that the frequency of the impulsive events, their 
 temporal and spatial distributions are consistent with  results  recorded by the  MWA \cite{Mondal-2020}. Furthermore, it has been also shown that the non- Poissonian feature as shown on Fig 7 of \cite{Mondal-2020}  is also naturally explained in the AQN framework. Indeed,     the AQN annihilation events  always will be accompanied by the clustering of radio events as the non-thermal electrons from one and the same AQN  may release their energy at different altitudes and different instants which lead to the clustering events as observed. 
 
 We conclude this section with the following comments. A close similarity  between the observed value for the  EUV luminosity $\sim 10^{27} {\rm erg~ s^{-1}}$ and computed value  within the  AQN framework is a highly nontrivial consequence  of the proposal  \cite{Zhitnitsky:2017rop}  as the general normalization   in the AQN based computation is determined by the DM density $\rho_{\rm DM}\approx 0.3~ \rm GeV/cm^3$ rather than by internal physics of the Sun. The emergence of a small  scale $\sim 200$ km which determines the TR is also entirely determined by the internal structure of the nuggets, their internal temperature and  ionization properties. Precisely these features determine very fast increase of the annihilation rate at the altitude around 2000 km as shown on Fig.\ref{luminosity}. Both these consequences of the AQN proposal can be considered as ``miracle coincidences" as there are no any free parameters in the corresponding estimates, see also footnote \ref{TR} with a relevant comment. 
 
 One may wonder if the  AQNs play any role in dynamics of the large solar flares which are characterized by  dramatically different energy scales (in comparison with nanoflares)  with  $W\simeq (10^{26}-10^{32}) ~\rm erg$.   It has been proposed  in \cite{Zhitnitsky:2018mav} that the AQNs hitting   the active regions with  large magnetic field  may   play the role of the triggers which could  ignite and initiate  the magnetic reconnections  leading to   large solar  flares. There is no room to elaborate on this interesting relation, and we refer to the original paper \cite{Zhitnitsky:2018mav}  for the details.

 The direct observation of the individual nanoflares which represent the AQN annihilation events according to the conjecture (\ref{W_B}) is hard to test directly   in the EUV or  X-ray regimes as the current instruments do not have sufficient resolution.  
 
At present time we think the most promising    tests of this proposal  can be achieved   in the radio frequency bands.  In particular, there must be  spatial  and temporal correlations between radio events in a given local region, in the  different frequency bands, with time delays measured in seconds. Similar  clustering events in the same frequency band have been observed by MWA \cite{Mondal-2020}.  We advocate the idea that similar spatial and temporal correlations must  also exist in  different frequency bands. This prediction can be directly tested  and analyzed  by  MWA since, according to 
\cite{Mondal-2020},  the data   in the 179, 196, 217 and 240 MHz bands have been  recorded, but not published yet. 

Another generic consequence of this framework is that the lower frequencies waves being emitted from higher altitudes must be suppressed while the intensity of the higher frequency bands must be enhanced. This is because the upward moving non-thermal electrons are much more numerous at lower altitude (corresponding to higher $\nu$) in comparison to higher altitudes (corresponding to lower $\nu$).   This prediction  can be also directly tested in  future studies. Finally, the Solar Orbiter recently observed so-called ``campfires"   in the  extreme UV frequency bands. It is tempting to identify such events with  the annihilation of large sized  AQN (which are rare events), as they are capable of generating radio signals sufficiently strong to be resolved. We therefore suggest to  search for a cross correlation between MWA radio signals and recordings of the extreme UV photons by Solar Orbiter.

  \section{The AQN model:   application to   observed ``Mysterious  Bursts" }\label{TA-bursts}
  In this section we review two recent papers  \cite{Zhitnitsky:2020shd,Liang:2021wjx}    devoted to explanation of the   observed by the Telescope Array (TA) puzzling events, the so-called  bursts\cite{Abbasi:2017rvx,Okuda_2019}. These events are very unusual and cannot be interpreted  in 
terms of  conventional single showers as reviewed below. These events have  been  coined  in \cite{Zhitnitsky:2020shd}  as the ``Mysterious  Bursts".

  We start in subsection \ref{TA-observations} by reviewing the unusual and very distinct  features of the TA-bursts \cite{Abbasi:2017rvx,Okuda_2019}, while in subsections \ref{AQN} and  \ref{TA-proposal} we explain how these puzzling   features   could be accommodated  within the AQN framework\cite{Zhitnitsky:2020shd}. Finally, in subsection \ref{TA-radio}  we argue that TA bursts will be inevitably accompanied by the  radio signals in  frequency band $\nu\in (0.5-200)$  MHz. These radio signals    must be synchronized with the TA bursts. The observation of such unique synchronization can unambiguously support, substantiate  or refute this proposal. 
  
  \subsection{The TA-bursts:  observations}\label{TA-observations}

  The unusual  features of the bursts recorded by  \cite{Abbasi:2017rvx,Okuda_2019} can be briefly formulated as follows:

1. {\it ``curvature puzzle".}   All reconstructed air shower fronts for the burst events are much more curved than usual cosmic ray (CR) air showers. This feature  is expressed in terms of the  time- spreading versus spatial-spreading of the particles in the bursts. The corresponding ``curvature" is much more pronounced for the bursts   in comparison with conventional CR air showers, see see Fig. 3 and Fig. 4 in \cite{Abbasi:2017rvx}. Furthermore, the bursts events do not have sharp edges in waveforms in comparison with  conventional CR events;

2. {\it ``clustering puzzle".} The events are temporally clustered within 1 ms, which would be a highly unlikely occurrence for three consecutive conventional high energy CR  hits  in the same area within a radius of approximately 1 km. The total 10 burst events have  been observed  during 5 years of observations.  
 \exclude{The authors of  ref. \cite{Abbasi:2017rvx} estimate the expectation of chance coincidence is less than $10^{-4}$ for five years of observations. If one tries to fit the rate of the observed bursts with conventional code, the energy for high energy CR events should be in $10^{13}$ eV energy range, while the observed intensity of the bursts correspond to $(10^{18}-10^{19})$ eV energy range.
 }
The estimated energy from individual events within the bursts is five to six orders of magnitude higher than the energy estimated by event rate. 

3. {\it ``synchronization puzzle".}  Most of the observed bursts are ``synchronized" (time delay between burst and lightning is less than 1 ms) or ``related" (time delay between burst and lightning is less than 200 ms) with the lightning events. Some of the bursts are not related to any lightnings. However, all 10 recorded bursts occur under thunderstorms.

It is very hard to understand all these features in terms of conventional CR physics as the bursts  cannot be reconciled  with conventional CR physics. 
At the same time   all unusual features (including the energetics, the flux estimates, the time and spatial spreading of each event within the bursts)    can be naturally explained  within the AQN framework. Before we proceed with corresponding explanation we have to briefly overview the basic features of the AQNs propagating under thunderclouds  to explain the nature and the source of the TA bursts.

 \subsection{The AQNs under the thunderclouds}\label{AQN}
The AQNs made of antimatter are capable to release a significant amount of  energy when they enter the Earth's atmosphere    and annihilation processes start to occur between  antimatter hidden in form of the AQNs and the   atmospheric  material. The emission of positrons from the nuggets made of antimatter  
during the thunderstorms  plays the  crucial role  in the proposal   \cite{Zhitnitsky:2020shd}.  The liberation of the positrons from the AQNs occurs  because  the thunderclouds are  characterized by large preexisting electric field $\cal{E}$.  This field liberates  and  accelerates   the  positrons which are normally bound to the AQNs with a typical average binding energy of order $\rm keV$. 
 
 The presence of the electric field $\cal{E}$ under thunderclouds is well established phenomenon. It  is  characterized by the following parameters \cite{Gurevich_2001,DWYER2014147}:
\be
 \label{parameters1}
 {\cal{E}}\simeq  {\rm \frac{kV}{cm}}, ~~~~  l_a \simeq 100 ~{\rm m},~~~~ \tau_{\cal{E}}\simeq \frac{l_a}{c}\simeq 0.3 {\rm\,\mu s},
 \ee  
  where $l_a$ is the so-called avalanche length.  
   If the AQN enters the  electric field (\ref{parameters1}) along  its path    it may liberate the positrons from the AQN's electrosphere 
  as the  additional energy $\Delta E$ assumes the same order of magnitude as the binding energy $E_{\rm bound} \sim \rm keV$    of the positrons, i.e.  
\be
\label{Delta_E}
\Delta E \simeq [e{\cal{E}}\cdot R_{\rm cap}]\sim 2 \rm ~keV\gtrsim E_{\rm bound},   
\ee
where parameter  $R_{\rm cap}$ is a typical distance   where positrons reside, and  can be estimated in terms of the ionization charge $Q$ and internal temperature of the nuggets $T\simeq 10 \rm ~keV$ as follows:
\be
\label{capture1}
R_{\rm cap}\simeq \frac{\alpha Q}{T}\sim 2~{\rm cm},
\ee
see  \cite{Zhitnitsky:2020shd} with proper estimates\footnote{\label{elastic}One should comment here that the positrons cannot be easily stripped away due to the elastic scattering with atmospheric material as the energy transfer $  E_{\rm tr} \sim m_ev_{\rm AQN}^2 $ 
 with $v_{\rm AQN}\sim 10^{-3}c$ measured in the   rest frame  of an AQN is not sufficient to liberate the positrons with $\rm keV$- binding energy.}. 
This additional energy (\ref{Delta_E})    of order of several  keV could  liberate the   positrons  from the nuggets, which consequently 
  will be accelerated   to MeV energies on  $l_a$  length scale. Indeed, 
 \be
\label{E_MeV}
  E_{\rm exit}\simeq [e{\cal{E}} \cdot l_a]\sim 10 \rm ~MeV. 
\ee
This estimate suggests that the positrons assume $10 \rm ~MeV$ energy after they   exit the region of strong  electric field which is known to exist  under thunderclouds.  It is  very important to emphasize that these estimates hold because the initial energy of the positrons is relatively high, in keV scale  according to (\ref{Delta_E}) such that they do not immediately annihilate, which would be the case for the positrons with eV energies. Due to the same reasons the positrons do not experience strong elastic scattering  and remain in the electric field background (\ref{parameters1}) during entire period of acceleration    which lasts about $0.3 {\rm\,\mu s}$, see also a footnote \ref{elastic} with related comment.

  \subsection{``Mysterious bursts" as   the AQN annihilation  events}\label{TA-proposal}
  The goal of this subsection is to explain   how the unusual features    listed   in subsection \ref{TA-observations} can be naturally understood within the AQN framework.

\subsubsection{``curvature puzzle"}
In the AQN proposal\cite{Zhitnitsky:2020shd}  the ``curved" feature can be easily understood by noticing that essential parameter in this proposal is the initial spread of the particles  determined by angle $ \Delta \alpha\simeq \left( {v_{\perp}}/{c}\right)\in (0-0.1)$. This spread in $\Delta \alpha$ is determined by   the velocity distribution   perpendicular to electric field
 at the exit point. The corresponding distribution   can be expressed in terms of the  initial energy (\ref{Delta_E})  as follows $v_{\perp}\simeq \sqrt{2 \Delta E/m}\lesssim 0.1c$.   
   Therefore, after travelling the distance $r$  the   spatially spread  range $\Delta s $ is estimated as 
\be
\label{spread}
\Delta s \simeq  r \left(\frac{\Delta \alpha}{\cos \alpha}\right) \simeq \frac{1~{\rm km}}{\cos\alpha} \left(\frac{r}{10 \rm ~km}\right)   \left(\frac{\Delta \alpha}{0.1}\right),   
\ee
  see Fig. \ref{geometry} for precise definitions of the parameters. 
  \exclude{
\begin{figure}
    \centering
    \includegraphics[width=0.8\linewidth]{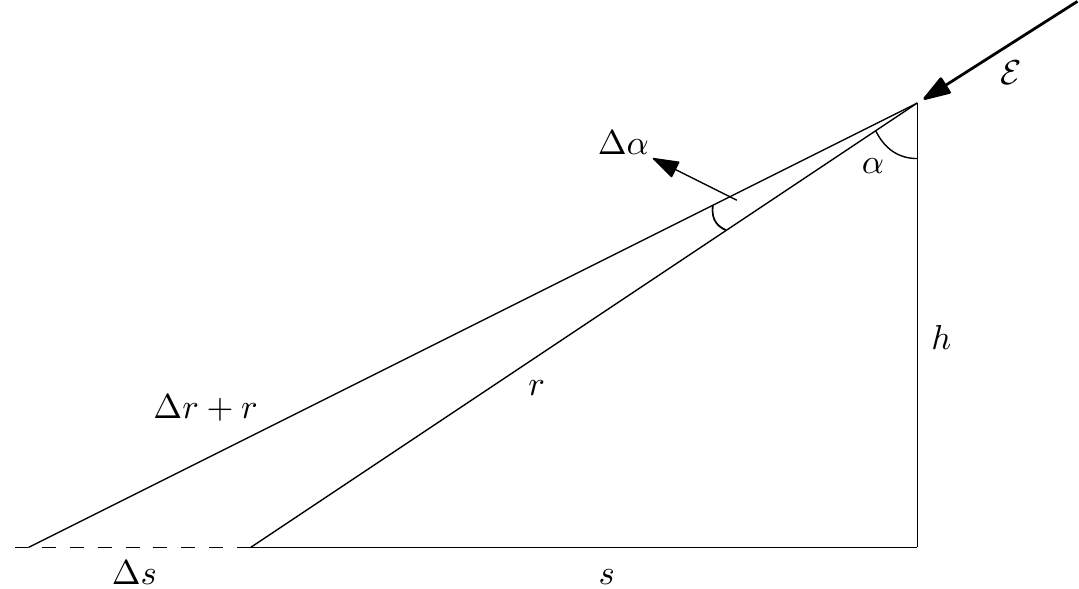}
    \caption{The positrons move along the cone  with angle $\Delta \alpha$ and  inclination angle $\alpha$ with respect to the vertical direction.  The angular spread  $\Delta \alpha\ll \alpha$ is assumed to be small. The spatial spread on the surface is determined  by $\Delta s$, while the additional travelling path is determined by $\Delta r$, see estimates in the text. The altitude is assumed to be within conventional range $ h\simeq (4-12)$ km. Instant  direction of the electric field $\mathbf{\cal{E}}$ at the moment of exit of the positrons   is also shown.}
    \label{geometry}
\end{figure}
}
At the same time, the temporal spread  $\Delta t$ can be estimated as follows:  
  \be
\label{spread1}
 \Delta t \simeq  \frac{\Delta r}{c}\simeq 3 {\rm\,\mu s} \cdot (\tan\alpha)\cdot 
   \left(\frac{r}{10 \rm ~km}\right)    \left(\frac{\Delta \alpha}{0.1}\right)
        \ee
       where $\Delta r \simeq  r \tan\alpha \Delta \alpha$,  
see Fig. \ref{geometry}. 
These parameters are linearly proportional to each other and assume  proper values     consistent with observations. Indeed,    
\be
\label{spread2}
 c\Delta t \simeq \Delta s \sin\alpha , ~~~\Delta \alpha\simeq \left(\frac{v_{\perp}}{c}\right)\in (0- 0.1)
\ee
such that $2\Delta t$ may vary between $(0.5-8) {\rm\,\mu s}$ when $2\Delta s$ changes between $(0.5-2)$ km with approximately linear slope determined by electric field direction $\sin \alpha$ which  is    consistent with observed events presented on Fig. 3 and Fig. 4 in \cite{Abbasi:2017rvx}. We use $(2\Delta t)$ and $2\Delta s$ in our estimates with extra factor 2 as  the angle $\Delta \alpha=(v_{\perp}/c)$        could assume the  positive or  negative value, depending on sign of $v_{\perp}$ with respect to instant direction of the electric field $\mathbf{{\cal{E}} }$ as shown on  
 Fig. \ref{geometry}.

 This behaviour in terms of the temporal  and spatial spreads  for the bursts  should be contrasted with conventional CR distribution when the timing spread  is much shorter and always below $2 {\rm\,\mu s}$ while the spatial spread is much longer, up to 3.5 km, see Fig. 5 in \cite{Abbasi:2017rvx}. This difference between the distributions in bursts and conventional CR air showers was coined as the ``curvature" puzzle, which    is naturally resolved within the AQN framework as argued above.  
 
 Similar  arguments also explain why the observed events do not demonstrate any  sharp edges in waveforms  (see Fig. 6 in  \cite{Abbasi:2017rvx}). This is because the conventional CR  air showers typically  have a single ultra-relativistic particle   generating  a very sharp edge in waveforms. It should be contrasted with large number of positrons  which  produce the non-sharp edges in waveforms in the  AQN-based proposal.  

  \begin{figure}
    \centering
    \includegraphics[width=4.0in]{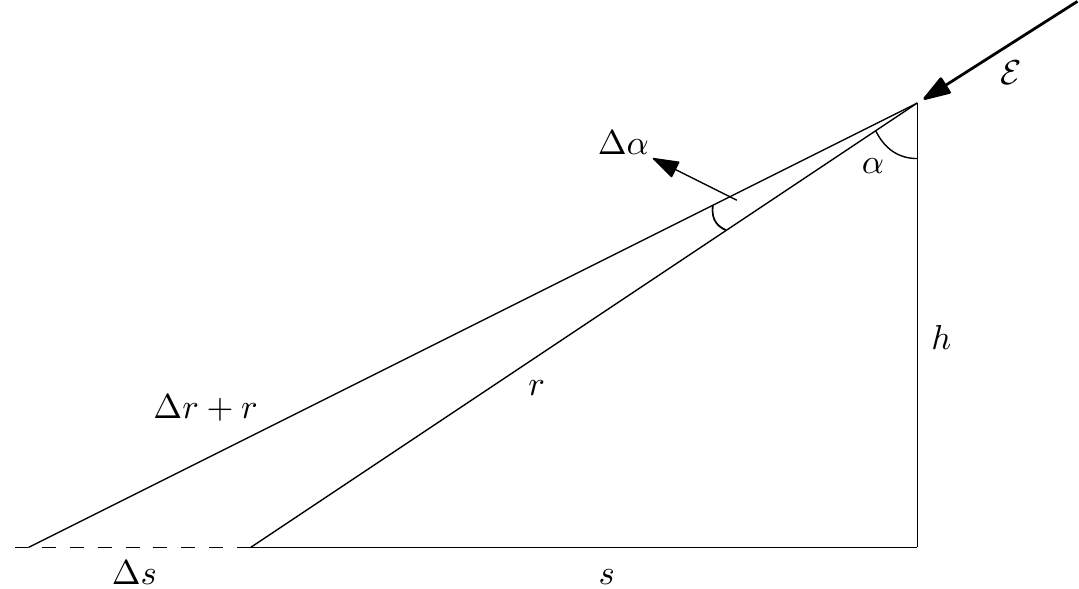}
    \caption{The geometry of the TA bursts, adopted from \cite{Zhitnitsky:2020shd}.   The angular spread of the propagating positrons is determined by $\Delta \alpha\ll \alpha$. The spatial spread on the surface is determined  by $\Delta s$, while the temporal spread  is determined by $\Delta r/c$. The altitude is assumed to be within conventional range $ h\simeq (4-12)$ km. Instant  direction of the electric field $\mathbf{\cal{E}}$ at the moment of exit of the positrons   is also shown.}
    \label{geometry}
\end{figure}

\subsubsection{``clustering puzzle"}
The ``clustering puzzle"  represents  a dramatic  inconsistency for the burst events if one assumes the randomness of the events described by conventional Poissonian distribution.   
The event rate suggests that the energy should be in $10^{13}$ eV  range, while the intensity of the events suggests  $(10^{18}-10^{19})$ eV   range, if one interpret the events as conventional CR air showers. It should be contrasted with the   AQN framework when the bursts represent  the {\it cluster}  of events (not a collection of independent random events).  The 
conventional Poissonian distribution does not apply here.   The presence of several events within the same burst is  a natural consequence of  the  
AQN's slow velocity when it merely propagates $\sim 0.25$ km during 1 ms and it always remains  in the same thunderstorm system characterized by the  fluctuating electric field $\cal{E}$.  The spatial spread for each individual   event within the same cluster also lies within  the range  (\ref{spread}),
being consistent with observations.

\subsubsection{``synchronization puzzle"  }

Most of the observed bursts are ``synchronized"  or ``related"   to the lightning events.  Few events   are not related to the lightnings, but  all 10 recorded bursts occurred under thunderstorms. This is very puzzling property  of the bursts if  interpreted  in terms of the conventional CR air showers because it is very hard to understand how the thunderstorms may dramatically modify CR features as discussed above. 
 
 At the same time  the ``synchronization" puzzle  is  perfectly consistent with AQN based proposal   \cite{Zhitnitsky:2020shd} 
because the thunderstorm with its pre-existing electric field (\ref{parameters1}) plays a crucial role in the  mechanism as the electric field 
is always present under the thunderclouds  irrespectively to the  lightning flashes. This strong electric field   instantaneously liberates the positrons and also accelerates them up to 10 MeV energies.   
  These positrons  can easily reach the TA surface detectors, and can produce the signals  consistent with  the bursts.

   \subsection{Radio signal from ``Mysterious bursts"}\label{TA-radio}
    In this subsection we highlight the basic ideas of the recent  studies \cite{Liang:2021wjx}  devoted to analysis of the radio signals which always accompany TA bursts when interpreted in terms of the AQN annihilation events under the thunderstorm as presented in previous section \ref{TA-proposal}. We shall argue below that the radio emission  is inevitable consequence of the proposed  explanation of the TA bursts when the positrons are accelerating in external electric field $\cal{E}$ under thunderclouds with typical duration   $\tau_{\cal{E}} \simeq 0.3 {\rm\,\mu s}$ as reviewed  in  subsection \ref{AQN}. 
     
 The starting point for these studies is spectral property of the electric field $\mathbf{E}_{\omega}$ at distance $R$,
 which is generated by  accelerating positrons:
    \be
 \label{spectral} 
 \mathbf{E}_{\omega}=\int_{-\infty}^{+\infty}\mathbf{E}e^{i\omega t}dt=\frac{e^{ikR}}{R} \left(\frac{Ne}{c^2}\right) \left(\frac{\omega}{\omega'}\right)^2\left[\mathbf{n}\times\left((\mathbf{n}-\frac{\mathbf{v}}{c})\times \mathbf{a}_{\omega'}\right)\right] ,~~~
  \ee 
 where all quantities at the right hand side of (\ref{spectral}) must be computed at the retarded times $t'\approx t-\frac{R}{c}+\frac{\mathbf{n}\cdot\mathbf{v}t'}{c}$.  The  coefficient  $N$ in formula (\ref{spectral}) is the number of the coherent positrons participating in the emission, while   parameters  $\omega'$ and  $\mathbf{a}_{\omega'}$ entering (\ref{spectral}) are defined as follows:
 \be
 \label{omega'}
 \omega'\equiv \omega \left(1-\frac{\mathbf{n}\cdot \mathbf{v}}{c}\right)\approx \frac{\omega}{2} \left[\frac{1}{\gamma^2}   + {\theta^2} \right], ~~\mathbf{a}_{\omega'}=\int_{-\infty}^{+\infty}\mathbf{a(t')}e^{i\omega' t'}dt', ~~\mathbf{a}(t)\approx\frac{e  {{\cal{E}}(t) }}{\gamma^3 m}.~~~
 \ee
The  spectral density of the emission $dE_{\omega}/d\omega$   can be computed in terms of these variables as follows:
        \be
 \label{angular-distribution1}
 \frac{dE_{\omega}}{ d\omega}&=&   \left(\frac{N^2 e^2\gamma^6 }{2\pi c^3}\right) \frac{16|\mathbf{a}_{\omega'}|^2}{ (1+\gamma^2\theta^2)^3}\left[ \frac{\gamma^2\theta^2}{1+\gamma^2\theta^2} \right]  \frac{d\Omega}{2\pi},~~~~\gamma\equiv\frac{1}{\sqrt{(1-\frac{v^2}{c^2})}},
  \ee 
  where $|\mathbf{a}_{\omega'}|$ should be  expressed in terms of $\theta$ according to (\ref{omega'}). Significance of this formula is that it explicitly shows that the emission mostly occurs along the small angles $\theta\simeq \gamma^{-1}$, as expected. Another important  feature of the 
   spectral density   $dE_{\omega}/d\omega$ is that the   radio pulse  occurs   in        bandwidth    $\nu\in (0.5-200) \rm ~MHz$ and lasts for about 
    $\tau_{\cal{E}} \simeq 0.3 {\rm\,\mu s}$, see  \cite{Liang:2021wjx}  for the details.
    
The next task  is to estimate the intensity of the  electric field 
 (\ref{spectral})  at very large distances $R$ where it can be potentially detected. The orientation of  $\mathbf{E}$  field 
 is determined by cross product  (\ref{spectral}) where one can    assume (for the simplicity of the numerical estimates) that $\mathbf{v}\parallel \mathbf{a}$.
 The absolute value for $|\mathbf{E}|$ at large distances  can be estimated as follows\footnote{One should not confuse a very small  electric field $|\mathbf{E}|$ of the radio pulse (\ref{E-field}) measured far away at distance $R$ from thunderclouds and very strong electric field $\cal{E}$ during the  lightnings  as given by (\ref{parameters1}) and measured in balloon experiments inside the thunderclouds.}
 \be
 \label{E-field}
 |\mathbf{E}| \approx  \frac{Ne  |\mathbf{a}|  \theta}{c^2R}\left(\frac{2\gamma^2}{1+\gamma^2\theta^2}\right)^3\approx
 90 {\rm  \frac{mV}{m}}    \left[\frac{(\gamma\theta)}{(1+\gamma^2\theta^2)^3} \right]
     \left(\frac{\gamma}{20 } \right)^2 \left(\frac{N}{10^9}\right)  \left(\frac{\rm 10 ~ km}{R}\right).~~
 \ee
 The distance $R$  in this equation should not be confused with parameter $r$ which enters all formulae from previous subsection \ref{TA-proposal}, including Fig. \ref{geometry}.  
   \begin{figure}
    \centering
    \includegraphics[width=4.0in]{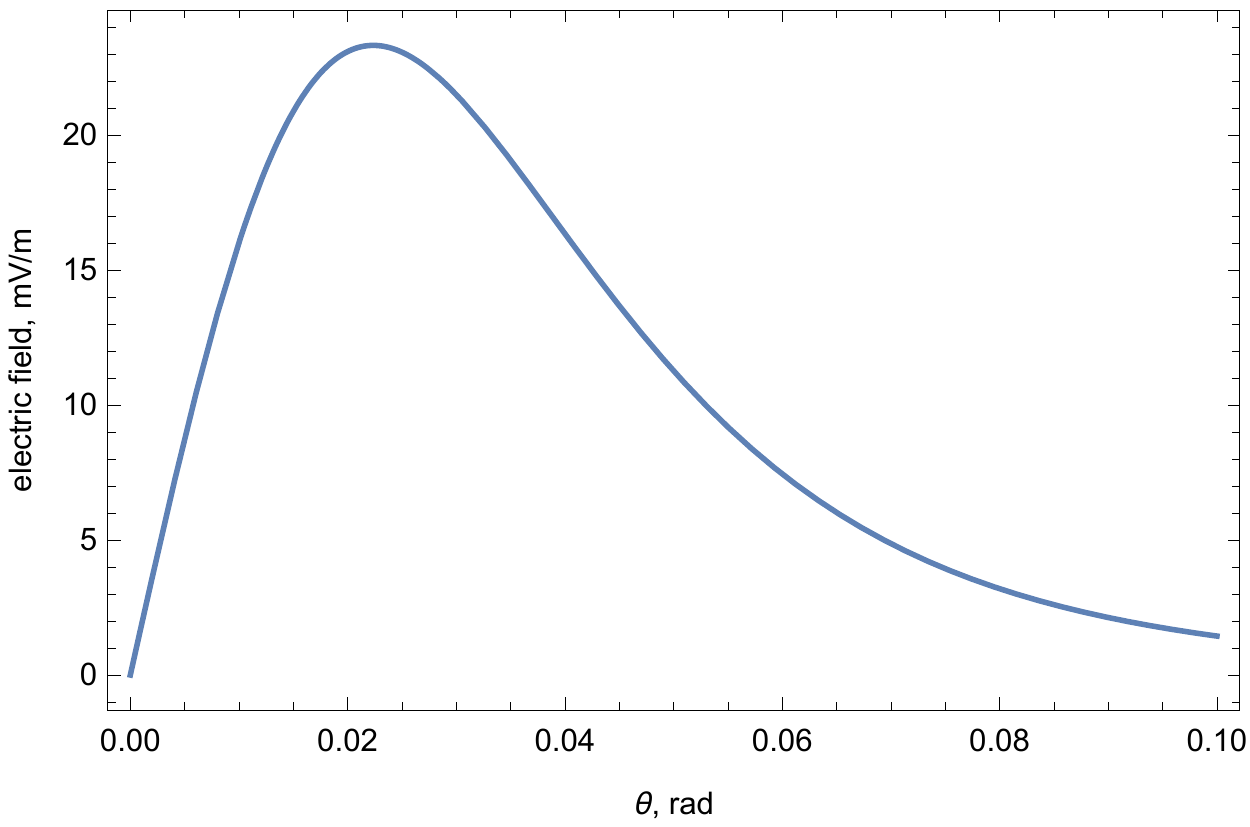}
    \caption{Strength of the AQN-induced electric field (\ref{E-field})  versus observation angle $\theta$, adopted from \cite{Liang:2021wjx}.  The parameters are chosen to be $N=10^9$, $\gamma=20$.  The electric field under thunderclouds is chosen to be ${\cal E}=1{\rm\,kV/cm}$ consistent with (\ref{parameters1}).}
    \label{Fig:Electric}
\end{figure}
 The expression (\ref{E-field})  has been derived under assumption that the acceleration $|\mathbf{a}|$  is a constant  
   during time $ t\in (0, \tau_{{\cal{E}}} )$. The dependence of $ |\mathbf{E} (\theta,t)|$ as a function of the $\theta$ is
    shown on Fig. \ref{Fig:Electric}   for typical parameters of the system. The key unknown parameter $N$  
   which enters   (\ref{E-field}) and which essentially determines the absolute value of the $ |\mathbf{E} (\theta,t)|$ field   was  estimated from  the assumption that the AQN induced positrons are   responsible for puzzling TA burst events with number of particles as recorded  by  \cite{Abbasi:2017rvx}. Therefore, the estimate (\ref{E-field}) is self-consistent with our  treatment of   the TA bursts as the AQN annihilation  events under the thunderstorm, as reviewed in subsection \ref{TA-proposal}. 
   
  One should emphasize that  the correlation between lightning and radio emission during thunderstorms   is well known and well documented   generic feature of the    lightning discharges. However, the AQN-induced radio pulses are qualitatively  distinct from conventional lightning-induced radio signals.  In particular, the lightning-induced radio emission is strongly peaked in few MHz bands, while AQN-induced radio pulse is characterized by the flat spectrum with $\nu\lesssim 200 \rm ~MHz$ according to \cite{Liang:2021wjx}. Furthermore, the AQN-induced radio pulses must    occur before the lightning flash or  at the  very initial moment of the lightning flashes as  the observed bursts demonstrate this feature \cite{Abbasi:2017rvx}.   It should be contrasted with the lightning-induced radio signals which  occur  during the     late stages of the lightning discharges. Based on these dramatic differences  in frequencies and   timings of the radio emissions  one  safely concludes  that the signals  due to the thunderstorm lightning events can be easily  discriminated from  the AQN-induced radio pulses.

 To summarize this section:  the  mysterious  bursts (with highly unusual features as  reviewed  in sections   \ref{TA-observations})     are naturally interpreted as the cluster events generated by the AQN annihilation events under  thunderstorm   as reviewed  in \ref{TA-proposal}.
 Some features of the system such as given by  (\ref{spread}), (\ref{spread1}), (\ref{spread2})
   are not sensitive to many  uncertainties related to complex dynamics of the AQN propagating under the thunderclouds. These features   represent almost model-independent consequences of the proposal   \cite{Zhitnitsky:2020shd} because they  are based exclusively on geometrical and kinematical features of the system.  
Furthermore, the radio pulse (\ref{E-field}) which is inevitable consequence of the proposal  must be  synchronized within $10 {\rm\,\mu s}$ with TA burst irrespectively whether the bursts are  related or unrelated    to the lightning events because  the positrons and radio waves emitted  from the same location at the same instant and both propagate with the speed of light.  Therefore, observing (not observing) such  synchronized   signals  can confirm, substantiate or refute this proposal.

  \section{The axions from AQNs: broadband axion searches}\label{axion}
  As we explained in sections \ref{introduction} and \ref{formation} the axion field is at heart of  the AQN construction  as it plays a dual role:  it makes the nuggets absolutely stable configurations as a result of extra pressure due to the axion domain walls surrounding the quark matter.
  The same $\cal{CP}$ odd axion field plays a vital role for baryon charge segregation, replacing the conventional baryogenesis.    However,  the corresponding axion energy (hidden in the form of the   axion domain wall) is not available unless  the AQN's baryon charges  from the nugget's core start annihilating processes with surrounding material. This  axion domain wall field which (in empty space)    equilibrates the Fermi pressure of the quark matter will start to adjust to these changes, and    the  axion energy  will be released into the space in the form of the  free propagating axions which can be observed by the axion detectors. 
  The goal of the present section is to highlight the basic ideas developed in \cite{Fischer:2018niu,Liang:2018ecs,Liang:2019lya,Budker:2019zka,Liang:2020mnz} on possible  strategy to search for such  AQN-induced axions.
 
  In next subsection \ref{axion-spectrum} we explain the dramatic difference between the galactic axions and the axions which are produced as a result of the annihilating  processes when the AQN enters the  atmosphere and crosses the Earth. The resulting spectrum of the axions will be drastically different from conventional galactic non-relativistic axions with $v_{\rm axion}\sim 10^{-3}c$. This difference in spectrum
  dictates a new  broadband strategy to search the AQN-induced axions which is the topic of the subsection \ref{axion-strategy}

  \subsection{Spectral features of the AQN-induced axions}\label{axion-spectrum}
   For the purposes of the present work it is sufficient  to mention  that the   conventional dark matter galactic  axions are  produced 
 due to the misalignment mechanism  when the cosmological field $\theta(t)$ oscillates and emits cold axions before it settles down at 
 its final destination $\theta_{\rm final}=0$, see recent reviews \cite{Marsh:2015xka,Irastorza:2018dyq,Sikivie:2020zpn}.
 Another mechanism is due  to the   decay of the topological objects.     It is important that  in both cases
 the produced  axions are non-relativistic particles with typical $v_{\rm axion}/c\sim 10^{-3}$, and their contribution to the dark matter density scales as $\Omega_{\rm axion}\sim m_a^{-7/6}$. This scaling  unambiguously implies that the axion mass must be fine-tuned  $m_a\lesssim 10^{-5}$ eV
 to  saturate the DM density today,  while larger axion mass will contribute very little to $\Omega_{\rm DM}$.
  The cavity type experiments have a potential to discover these  non-relativistic axions. 
 Axions can be also produced as a result of the Primakoff effect in a stellar plasma at high temperature see recent reviews  \cite{Marsh:2015xka,Graham:2015ouw,Irastorza:2018dyq,Sikivie:2020zpn}. These axions are ultra-relativistic as the typical average energy of the axions emitted by the Sun is $\la E\ra =4.2$ keV. 
 
  There is a fundamentally novel mechanism of the  axion production  when the AQNs enter stars or planets\cite{Fischer:2018niu}. 
This mechanism is rooted to  the  AQN dark matter model.   The most important feature of the emitted axions     is that these emitted axions   will be released with relativistic (but not ultra-relativistic) velocities  with 
typical values $v_{\rm axion}^{\rm AQN}\simeq 0.6 c$. These features should be   contrasted with conventional galactic non-relativistic axions    $v_{\rm axion}\sim 10^{-3}c$ and solar ultra-relativistic axions with typical energies 
$\la E\ra =4.2$ keV. 
\begin{figure}
    \centering
    \includegraphics[width=4.0in]{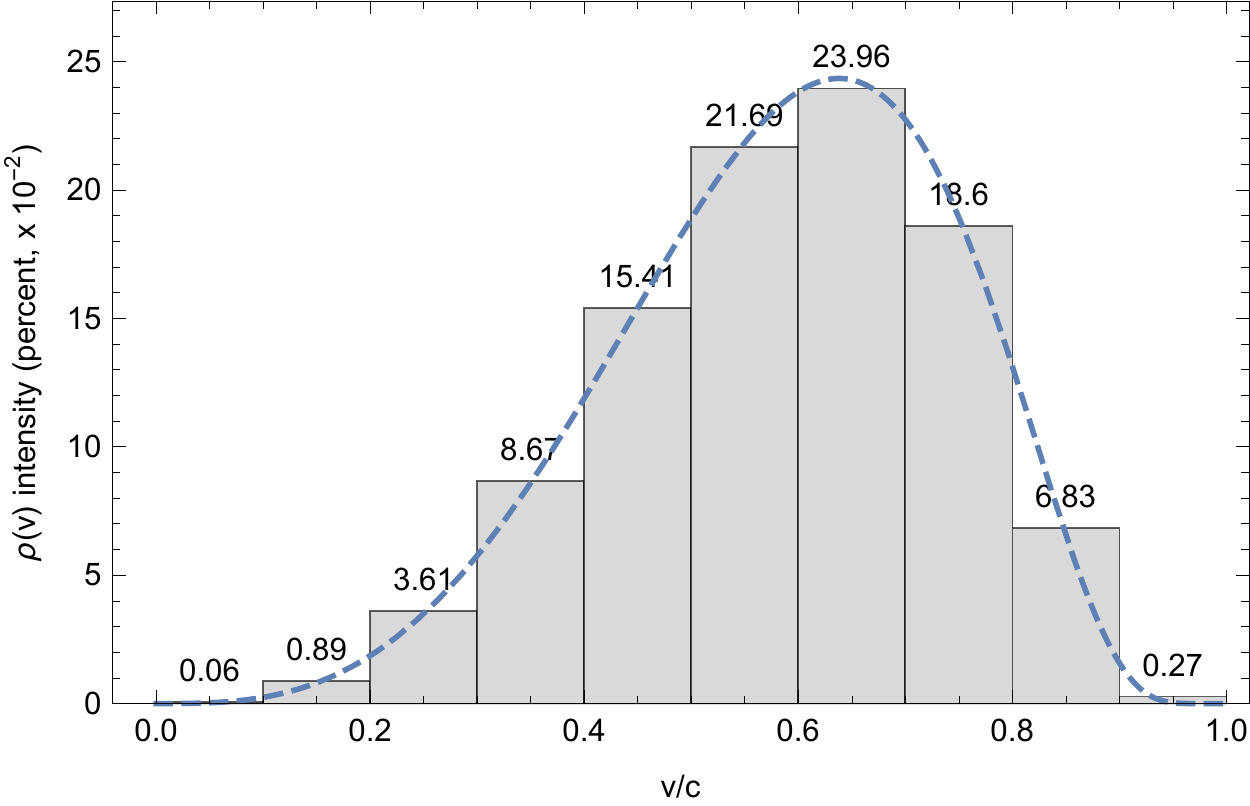}
    \caption{Normalized spectrum $\rho(v)$ of the AQN-induced axions, adopted from \cite{Liang:2018ecs}.}
    \label{fig:spectrum}
\end{figure}

The new mechanism of production of these axions can be explained as follows. The total energy of an AQN finds its equilibrium minimum when the axion domain wall contributes about 1/3 of its total mass  \cite{Ge:2017idw}. 
This  configuration  in the equilibrium does not emit any axions as a result of pure kinematical constraint: the static domain wall axions are off-shell non-propagating axions.   However, this static picture    drastically changes when some baryon charge from the AQN get annihilated as a result of interaction with environment when  the time dependent perturbations  obviously change this equilibrium configuration.  
In other words,  the  configuration becomes    unstable with respect to emission of the axions because the AQN  is no longer at its minimum energy configuration with fewer baryon charge in the quark nugget core. 
 As a result of these annihilation processes, the AQN starts to loose  its mass, and consequentially its size starts to shrink. To reiterate: the  annihilation of antinuggets when an AQN hits the stars or planets  forces the surrounding domain wall to oscillate.  These oscillations of domain wall generate excitation modes and ultimately lead to radiation of the propagating axions. The spectrum of the corresponding 
AQN-induced axions has been computed in \cite{Liang:2018ecs}, and we present the corresponding results in Fig. \ref{fig:spectrum}.

There are several important features of this spectrum which are deserved to be mentioned here. First of all, typical value $v_{\rm axion}^{\rm AQN}\simeq 0.6 c$ is very large in comparison with velocities of   conventional galactic non-relativistic axions    $v_{\rm axion}\sim 10^{-3}c$. When the AQN hits the Earth and the annihilation processes start  the axions will be also emitted   as explained above.  The corresponding axion density on the Earth's surface has been computed using full scale Monte Carlo simulations \cite{Liang:2019lya}. The   results of these computations  can be   expressed as follows:  
\be
\label{denisty_relativitic}
\la \rho_\mathrm{a}^{\rm AQN}(R_\oplus)\ra \sim 5\cdot 10^{-6}{\rm  \frac{GeV}{cm^3}}, ~~~ \la v_\mathrm{a}^{\rm AQN}(R_\oplus)\ra \simeq 0.6 c.
\ee
These axions are mostly produced in  deep Earth's underground where the density of surrounding material is the highest.   One should also note that on average a typical nugget looses approximately $30\%$ of its baryon charge when crosses the Earth. Axion domain wall shrinks correspondingly, which eventually generates the axion density (\ref{denisty_relativitic}). The resulting  number  density $n_a^{\rm AQN}\simeq \rho_\mathrm{a}^{\rm AQN}/m_a$  is approximately 5 orders of magnitude smaller than conventional galactic axion number density assuming that the galactic non-relativistic axions saturate the DM density observed today. However, the flux $(v_\mathrm{a}^{\rm AQN} n_a^{\rm AQN})$ related to   relativistic axions (\ref{denisty_relativitic}) is only 2 orders of magnitude smaller than conventional flux of non-relativistic axion. Furthermore, due to the greater  velocities
$v_\mathrm{a}^{\rm AQN} $ these axions interact with material in dramatically different way, which could strongly enhance the likelihood of their detection. 

One should also mention that these axions   can be treated as a classical field because the number   of the AQN-induced axions (\ref{denisty_relativitic}) accommodated by a single de-Broglie volume is very large in spite of the fact that the de-Broglie wavelength $\lambda$ for relativistic AQN-induced axions  is much shorter than for galactic axions, 
 \be
\label{density}
n_a^{\rm AQN}\lambda^3\sim \frac{\la \rho_\mathrm{a}^{\rm AQN}(R_\oplus)\ra}{m_a}\cdot \left(\frac{\hbar}{m_a v_a}\right)^3\sim 10^6 \left(\frac{10^{-4} {\rm eV}}{m_a}\right)^4\gg 1. \nonumber
 \ee

Another comment we would like to make is as follows. The production of the low energy axions with $v_a\ll c$ is strongly suppressed as one can see from  Fig. \ref{fig:spectrum}. However, the axions which are produced with  extremely  low velocities $v_a\lesssim 11 \rm ~ {km}/{s}$ will be   trapped  by  the Earth's gravitational field.  These axions will be orbiting  the Earth indefinitely, and therefore they will be accumulated around the Earth during entire life time which is 4.5 billion years.  
 
The corresponding  Monte Carlo simulations have been performed in \cite{Liang:2018ecs} with the following estimate:
\be
\label{denisty_bound}
\rho_\mathrm{a}(R_\oplus)\sim 10^{-4}{\rm  \frac{GeV}{cm^3}}, ~~~ \la v_\mathrm{a}(R_\oplus)\ra \simeq 8 \rm \frac{km}{s}~~~ [\rm gravitationally ~ bound~ axions].
\ee
The number density   of the bound axions is at least 2 orders of magnitude smaller than  conventional axion number density assuming that the galactic non-relativistic axions saturate the DM density. However, the corresponding wavelength $\lambda_\mathrm{a}\sim {\hbar}/({m_\mathrm{a} v_\mathrm{a}})$ of the gravitationally bound axions is approximately $30$ times greater than for galactic axions, which have a typical velocity of about $\sim 250  $ km/s. Therefore, coherent effects can be maintained for a longer period of  time  in comparison with the case of  conventional galactic axion searches. One may hope that this feature of having a large coherence length, $\lambda_\mathrm{a}\sim v_\mathrm{a}^{-1}$, could play a key role in design of specific instruments, capable of discovering such gravitationally trapped axions. 
  
   \subsection{Broadband search strategy  for the AQN-induced axions}\label{axion-strategy}
 The large average velocities   $\la v_a\ra \simeq 0.6 c$ of the emitted axions by AQNs  dramatically changes   entire  strategy of axion searches. This is because the axions are characterized by broad distribution with $m_a\lesssim \omega_a\lesssim 1.8~m_a$ as discussed in previous subsection.   Therefore, the corresponding axion detectors  
   must be some kind of broadband instruments. For example, if $m_a\approx 1$GHz, the detectors must be sensitive at least to the window $(1-2) $ GHz. 
   The cavity type experiments such as ADMX are to date the only ones to probe the parameter space of the conventional QCD axions with $\la v_a\ra \sim 10^{-3} c$, while we are interested in detection of the relativistic axions with  $\la v_a\ra \sim 0.6 c$. This requires a different type of instruments and drastically different search strategies.  We assume that some 
    kind of broadband instruments can be designed and built, see reviews 
     \cite{Marsh:2015xka,Graham:2015ouw,Irastorza:2018dyq,Sikivie:2020zpn} with description of   possible  detectors. 
   
     With this assumption in mind, a strategy to probe the QCD axion can be formulated as follows\cite{Budker:2019zka}. It has been known since \cite{Freese:1987wu}   that the DM flux shows  annual modulation due to the differences in  relative orientations  of the DM wind and the direction of the Earth motion around the Sun, which generates the flux difference.  The corresponding effect for AQN induced  axions was
 computed in  \cite{Liang:2019lya}.   The daily modulations have been largely   ignored in the past because they are numerically very small for WIMP like models.
 Indeed, the velocity difference due to the Earth's rotation about its axis is only $\sim$ 0.5 km/s, to be compared with galactic wind $\sim 250  $ km/s. The daily modulation in the AQN model  is   a  very specific  feature of the AQN model. Full scale Monte Carlo simulations carried out in \cite{Liang:2019lya} have shown that the daily modulation could be very large\footnote{\label{daily}The   daily modulations in the AQN model is a consequence of the AQN's size difference between the moment of entry and moment of exit resulting  from annihilation processes during the propagation  in the Earth's interior. Such effects  do no exist for any  fundamental particles such as WIMPs. This difference  may generate a large effect $\sim 10\%$   for the daily modulations, see \cite{Liang:2019lya} for the details.}.

The broadband strategy is to separate a large frequency  band into a number of smaller frequency bins  with the width $\Delta \nu \sim \nu$ according to the axion dispersion relation as discussed above. 
 The time dependent signal in each frequency bin $\Delta \nu_i$ has to be fitted according to the expected modulation pattern, daily, or annual. For example, the annual modulation should be fitted according to the following formula 
  \begin{equation}
\label{eq:annual}
 A_{\rm (a)}(t)\equiv[1+\kappa_{\rm (a)} \cos\Omega_a (t-t_0)],~~~~  \rho_\mathrm{a}^{\rm AQN}(t) \equiv A_{\rm (a)}(t) \la \rho_\mathrm{a}^{\rm AQN}(R_\oplus)\ra
\end{equation}
 where $\Omega_a=2\pi\,\rm yr^{-1} $ is the angular frequency of the annual modulation
 and   label $``a"$ in $\Omega_a $ stands for annual.   The $\Omega_a t_0$ is the phase shift corresponding to the maximum on June 1 and minimum on December 1 for the standard galactic DM distribution, see \cite{Freese:1987wu,Freese:2012xd}. 
 
 The same procedure should be repeated for all frequency bins ``$i$".
 Let us assume that the modulation has been recorded in a specific bin $\bar{i}$. 
 The modulation coefficient $\kappa^{\bar{i}}_{\rm (a)}$ for a specific  $\bar{i}$ could be as large as $10\%$. The parameters $\Omega_a$, $\kappa^{\bar{i}}_{\rm (a)}$ and $t_0$ are to be extracted from the fitting analysis and compared with theoretical predictions. 
 
 A test that it is not a spurious signal is a relatively simple procedure: one should check that no modulations appear in all other bins (except to possible neighbours to  $\bar{i}$ bin). Furthermore, one should also check that no modulation occurs for zero magnetic field when the axion 
 -photon conversion cannot arise.   
   
 A similar procedure can be applied for the daily modulations and  can be expressed  as follows: 
 \begin{equation}
\label{eq:daily}
A_{\rm (d)}(t)\equiv[1+\kappa_{\rm (d)} \cos(\Omega_d t-\phi_0)],~~~~ \rho_\mathrm{a}^{\rm AQN}(t) \equiv A_{\rm (d)}(t) \la \rho_\mathrm{a}^{\rm AQN}(R_\oplus)\ra
\end{equation}
 where   $\Omega_d=2\pi\,\rm day^{-1} $ is the angular frequency of the daily modulation, while $\phi_0$ is the phase shift similar to $\Omega_at_0$ in (\ref{eq:annual}). It   can be assumed to be constant on the scale of days. However, it actually slowly changes  during the annual seasons due to the  variation of the direction of  DM wind   with respect  to the Earth, see footnote \ref{daily} on the nature of the daily modulations. This feature can be used  as a  test  to remove the noise as the phase $\phi_0$ should  change by $\pi$ in 1/2 year  such that the daily modulations flip the sign in  1/2 year.
 
 The daily modulations are much easier to analyze than annual modulations because it obviously requires less time to collect sufficient statistics\footnote{For example, accumulation the data during 3 months (90 days)  when the phase $\phi_0$ remains approximately constant, may give us some hints on daily modulations (\ref{eq:daily}) as  90 complete cycles are being accumulated. The same period of time  is obviously too short   to observe the annual modulation  given by  (\ref{eq:annual}) as a single  annual cycle is far from being complete.}. Furthermore, the daily modulation is very specific feature of the AQN framework which is not shared by conventional WIMP-like DM candidates, see footnote \ref{daily} with a comment. Therefore, the recording of the daily modulations  leading to non-vanishing  $\kappa_{\rm (d)}$ on the $10\%$ level would be a very strong support for the AQN model. One should also add that any axion search instruments presently operating can, in principle, analyze  the daily modulations along the lines described above. This obviously may include all previous data sets. Such studies can be carried out by any  haloscope irrespectively to the conventional searches based on resonance scanning  as the basic idea is to combine entire data set (let us say collected during  a specific  month when phase $\phi_0$ from (\ref{eq:daily}) can be assumed to be constant) for each given hour to see if the data show any daily modulation  for a specific frequency band of the haloscope.  
 
 In summary, the AQN-induced axions characterized by broad distribution with $m_a\lesssim \omega_a\lesssim 1.8~m_a$ as discussed above  
    will produce nonzero modulation coefficients $\kappa_{\rm (a)}$ and $\kappa_{\rm (d)}$ in one frequency  bin  $\bar{i}$ (or perhaps two neighbouring bins\footnote{For example, for  $m_a \simeq 1.25\cdot 10^{-5} \rm eV$ the bin width is very large: $\Delta \nu\in (3, 5.4) \rm GHz $  which is precise manifestation of the broad band requirement  as discussed above. It should be contrasted with conventional cavity type experiments when $\Delta \nu/\nu\sim 10^{-6}$. }). It is a nontrivial consistency test that the modulation occurs in one and the same frequency bin $\bar{i}$ for two drastically different analyses: the fittings for (\ref{eq:annual}) and (\ref{eq:daily}), correspondingly. A further consistency check is see whether the modulation is observed in other frequency bins. Another consistency check is  the zero field test, as we already mentioned. One more consistency check  is the study of  the phase $\phi_0$ which must demonstrate the drift with a season.  This is also important test to remove the spurious signals, 
    even when the axion   detectors are not designed   for the broadband searches. 
     Finally,  a more powerful  test to exclude a spurious signal is  based on idea to use some kind of network of synchronized instruments to study correlated   signals. It should be considered as a unique  tool which discriminates the  true signal contributing to  (\ref{eq:annual}) and (\ref{eq:daily}) from a spurious noise background.  This topic  will not be covered by this review, and we refer to the original papers  \cite{Budker:2019zka,Liang:2020mnz} for the details.

    \section{Conclusion}\label{conclusion}
    We conclude this short review with the following comments.
    The AQN framework was initially invented to explain in a very natural way the observed similarity   
between visible and dark components   of  the Universe: $\Omega_{\rm DM} \approx   \Omega_{\rm B}$. This generic relation is a direct consequence of the construction and does not depend on any specific parameters of the model as  both components are proportional to the same fundamental $\Lambda_{\rm QCD} $ scale,  
and   both components are originated at the same  QCD epoch as reviewed in sections \ref{introduction} and \ref{formation}.

 The construction inevitably includes the antimatter in CS phase as a part of the dark component. The anti-quarks   are not easily available for annihilation unless the AQNs hit the stars and planets,  though  some rare events of annihilations at the galactic center (where both components, the visible and dark, are sufficiently high) can also occur.  We reviewed three specific recent applications of this framework:  in Section \ref{corona}  we overviewed  a possible resolution  of the  ``solar corona  mystery" while in Section \ref{TA-bursts}  we highlighted  a   possible resolution  of the ``mysterious TA bursts". We also reviewed in Section \ref{axion} a new broadband strategy to discover the AQN-induced axions which are at the heart of the construction.  Each  section devoted to a specific  application concluded with a short summary where   a number of independent experiments, tests or observations is suggested to  confirm, substantiate or refute this proposal. There is no need to repeat these summaries again in Conclusion. 
 
 Instead, I would like to mention several other directions for future  studies which had not been covered by this review  due to the size limitation. 
In particular,  the AQN model may  explain  some observed excesses of diffuse emission from the galactic center the origin of which  remains to be debated,   see the original works \cite{Oaknin:2004mn, Zhitnitsky:2006tu,Forbes:2006ba, Lawson:2007kp,Forbes:2008uf,Forbes:2009wg} with explicit  computations of the galactic radiation  excesses  for varies frequencies, including excesses of the diffuse  X- and   $\gamma$- rays.  
In all these cases photon emission is originated  from   the electrosphere, and all intensities in different frequency bands are expressed in terms of a single parameter $\langle B\rangle $ entering formula  (\ref{eq:observable}).  Future observations, including the studies of the intensity and morphology of the well known 511 keV line may finally shed some  light on the source  of the   excess  of radiation, which is still under active debates. 

The  AQNs may also offer a resolution of   the     ``Primordial Lithium Puzzle"  as suggested in \cite{Flambaum:2018ohm}.  
\exclude{Another  longstanding puzzle related to  the DAMA/LIBRA  observation  of the annual modulation at $9.5\sigma$ confidence level  can be also resolved within the same AQN framework as argued in \cite{Zhitnitsky:2019tbh}. 
This resolution can be confirmed or refuted by independent studies by COSINE-100, CYGNO or ANAIS-112 collaborations as described in \cite{Zhitnitsky:2019tbh}. 
}
The  AQNs may also resolve the observed  (by XMM-Newton at  $ 11\sigma$ confidence level  \cite{Fraser:2014wja})  puzzling   seasonal variation   of the X-ray background in the near-Earth environment in the 2-6 keV energy range as suggested in  \cite{Ge:2020cho}. 
   The AQN annihilation events in the Earth's atmosphere could produce  infrasound and seismic acoustic waves and one can study these effects using the Distributed Acoustic Sensors   or  modern seismometers  as suggested in   \cite{Budker:2020mqk,Figueroa:2021bab}. In fact, it has been  further speculated in  \cite{Budker:2020mqk} that a mysterious explosion which occurred on July 31st 2008 and which was properly recorded by the dedicated Elginfield Infrasound Array
 might be a good candidate for an AQN-annihilation event with very large $B\simeq 10^{27}$  as the basic estimates for the overpressure $\delta p\approx 0.3$ Pa
    and the infrasound frequency $\nu \sim 5$ Hz are amazingly close to the recorded signal.  It has been also argued that two anomalous events     with noninverted polarity as observed by Antarctic Impulse Transient Antenna (\textsc{ANITA})  collaboration \cite{Gorham:2016zah,Gorham:2018ydl}  could be explained within the same AQN framework with the same fundamental parameters  \cite{Liang:2021rnv}. 
    These events are proven  to be hard  to explain in terms of conventional cosmic rays, while the AQN framework offers a natural explanation without introducing any additional parameters.   
   This list is already very long, but  obviously far from being complete.
    We conclude on this optimistic note.

\section*{Acknowledgments}

I am thankful to all my  co-authors  from variety  of fields (particle physics experiment and theory,   nuclear physics, Atomic, Molecular, Optic (AMO) physics, astronomy)    who enormously contributed to development of the AQN framework. 
It would not be possible to make a progress in this very broad  project  without such fruitful  collaborations. 
This research was supported in part by the Natural Sciences and Engineering
Research Council of Canada.

\bibliographystyle{ws-mpla}
\bibliography{review}

\begin{thebibliography}{10}

\bibitem{Sakharov:1967dj}
A.~Sakharov, {\em JETP Lett.} {\bf 5}, 24  (1967).

\bibitem{Witten:1984rs}
E.~{Witten}, {\em \prd} {\bf 30}, 272 (July 1984).

\bibitem{Farhi:1984qu}
E.~{Farhi} and R.~L. {Jaffe}, {\em \prd} {\bf 30}, 2379 (December 1984).

\bibitem{DeRujula:1984axn}
A.~{De Rujula} and S.~L. {Glashow}, {\em \nat} {\bf 312}, 734 (December 1984).

\bibitem{Zhitnitsky:2002qa}
A.~R. {Zhitnitsky}, {\em \jcap} {\bf 10},   010 (October 2003),
  \href{http://arxiv.org/abs/hep-ph/0202161}{{\ttfamily hep-ph/0202161}}.

\bibitem{Zhitnitsky:2006vt}
A.~{Zhitnitsky}, {\em \prd} {\bf 74},   043515 (August 2006),
  \href{http://arxiv.org/abs/astro-ph/0603064}{{\ttfamily astro-ph/0603064}}.

\bibitem{Ge:2019voa}
S.~Ge, K.~Lawson and A.~Zhitnitsky, {\em Phys. Rev. D} {\bf 99},   116017
  (2019), \href{http://arxiv.org/abs/1903.05090}{{\ttfamily arXiv:1903.05090
  [hep-ph]}}.

\bibitem{Flambaum:2018ohm}
V.~V. Flambaum and A.~R. Zhitnitsky, {\em Phys. Rev. D} {\bf 99},   023517
  (2019), \href{http://arxiv.org/abs/1811.01965}{{\ttfamily arXiv:1811.01965
  [hep-ph]}}.

\bibitem{Liang:2016tqc}
X.~{Liang} and A.~{Zhitnitsky}, {\em \prd} {\bf 94},   083502 (October 2016),
  \href{http://arxiv.org/abs/1606.00435}{{\ttfamily arXiv:1606.00435
  [hep-ph]}}.

\bibitem{Ge:2017ttc}
S.~{Ge}, X.~{Liang} and A.~{Zhitnitsky}, {\em \prd} {\bf 96},   063514
  (September 2017), \href{http://arxiv.org/abs/1702.04354}{{\ttfamily
  arXiv:1702.04354 [hep-ph]}}.

\bibitem{Ge:2017idw}
S.~{Ge}, X.~{Liang} and A.~{Zhitnitsky}, {\em \prd} {\bf 97},   043008
  (February 2018), \href{http://arxiv.org/abs/1711.06271}{{\ttfamily
  arXiv:1711.06271 [hep-ph]}}.

\bibitem{1977PhRvD..16.1791P}
R.~D. {Peccei} and H.~R. {Quinn}, {\em \prd} {\bf 16}, 1791 (September 1977).

\bibitem{1978PhRvL..40..223W}
S.~{Weinberg}, {\em Phys. Rev. Lett.} {\bf 40}, 223 (January 1978).

\bibitem{1978PhRvL..40..279W}
F.~{Wilczek}, {\em Phys. Rev. Lett.} {\bf 40}, 279 (January 1978).

\bibitem{KSVZ1}
J.~E. {Kim}, {\em Phys. Rev. Lett.} {\bf 43}, 103 (July 1979).

\bibitem{KSVZ2}
M.~A. {Shifman}, A.~I. {Vainshtein} and V.~I. {Zakharov}, {\em Nucl. Phys. B}
  {\bf 166}, 493 (April 1980).

\bibitem{DFSZ1}
M.~{Dine}, W.~{Fischler} and M.~{Srednicki}, {\em Phys. Lett. B} {\bf 104}, 199
  (August 1981).

\bibitem{DFSZ2}
A.~R. Zhitnitsky, {\em Sov. J. Nucl. Phys.} {\bf 31},   260  (1980), [Yad.
  Fiz.31,497(1980)].

\bibitem{Marsh:2015xka}
D.~J.~E. {Marsh}, {\em Phys. Rep.} {\bf 643}, 1 (July 2016),
  \href{http://arxiv.org/abs/1510.07633}{{\ttfamily arXiv:1510.07633}}.

\bibitem{Graham:2015ouw}
P.~W. {Graham}, I.~G. {Irastorza}, S.~K. {Lamoreaux}, A.~{Lindner} and K.~A.
  {van Bibber}, {\em Ann. Rev. of Nucl. and Part. Sc.} {\bf 65}, 485 (October
  2015), \href{http://arxiv.org/abs/1602.00039}{{\ttfamily arXiv:1602.00039
  [hep-ex]}}.

\bibitem{Irastorza:2018dyq}
I.~G. Irastorza and J.~Redondo, {\em Prog. Part. Nucl. Phys.} {\bf 102}, 89
  (2018), \href{http://arxiv.org/abs/1801.08127}{{\ttfamily arXiv:1801.08127
  [hep-ph]}}.

\bibitem{Sikivie:2020zpn}
P.~Sikivie, {\em Rev. Mod. Phys.} {\bf 93},   015004  (2021),
  \href{http://arxiv.org/abs/2003.02206}{{\ttfamily arXiv:2003.02206
  [hep-ph]}}.

\bibitem{Vilenkin1994}
A.~Vilenkin and E.~Shellard, {\em Cosmic strings and other topological defects}
    (Cambridge University Press, 1994).

\bibitem{Kolb:1990vq}
E.~W. Kolb and M.~S. Turner, {\em {The Early Universe}} 1990.

\bibitem{Lawson:2019cvy}
K.~Lawson, X.~Liang, A.~Mead, M.~S.~R. Siddiqui, L.~Van~Waerbeke and
  A.~Zhitnitsky, {\em Phys. Rev. D} {\bf 100},   043531  (2019),
  \href{http://arxiv.org/abs/1905.00022}{{\ttfamily arXiv:1905.00022
  [astro-ph.CO]}}.

\bibitem{Gorham:2012hy}
P.~Gorham, {\em Phys. Rev.} {\bf D86},   123005  (2012),
  \href{http://arxiv.org/abs/1208.3697}{{\ttfamily arXiv:1208.3697
  [astro-ph.CO]}}.

\bibitem{Jacobs:2014yca}
D.~M. {Jacobs}, G.~D. {Starkman} and B.~W. {Lynn}, {\em \mnras} {\bf 450}, 3418
  (July 2015), \href{http://arxiv.org/abs/1410.2236}{{\ttfamily
  arXiv:1410.2236}}.

\bibitem{Zhitnitsky:2017rop}
A.~{Zhitnitsky}, {\em \jcap} {\bf 10},   050 (October 2017),
  \href{http://arxiv.org/abs/1707.03400}{{\ttfamily arXiv:1707.03400
  [astro-ph.SR]}}.

\bibitem{Zhitnitsky:2018mav}
A.~Zhitnitsky, {\em Phys. Dark Univ.} {\bf 22}, 1  (2018),
  \href{http://arxiv.org/abs/1801.01509}{{\ttfamily arXiv:1801.01509
  [astro-ph.SR]}}.

\bibitem{Raza:2018gpb}
N.~Raza, L.~van Waerbeke and A.~Zhitnitsky, {\em Phys. Rev. D} {\bf 98},
  103527  (2018), \href{http://arxiv.org/abs/1805.01897}{{\ttfamily
  arXiv:1805.01897 [astro-ph.SR]}}.

\bibitem{Ge:2020xvf}
S.~Ge, M.~S.~R. Siddiqui, L.~Van~Waerbeke and A.~Zhitnitsky, {\em Phys. Rev.}
  {\bf D102},   123021  (2020),
  \href{http://arxiv.org/abs/2009.00004}{{\ttfamily arXiv:2009.00004
  [astro-ph.HE]}}.

\bibitem{Parker}
E.~N. {Parker}, {\em \apj} {\bf 330}, 474 (July 1988).

\bibitem{Benz-2000}
S.~{Krucker} and A.~O. {Benz}, {\em \solphys} {\bf 191}, 341 (February 2000),
  \href{http://arxiv.org/abs/astro-ph/9912501}{{\ttfamily astro-ph/9912501}}.

\bibitem{Kraev-2001}
U.~{Mitra-Kraev} and A.~O. {Benz}, {\em \aap} {\bf 373}, 318 (July 2001),
  \href{http://arxiv.org/abs/astro-ph/0104218}{{\ttfamily astro-ph/0104218}}.

\bibitem{Benz-2002}
A.~O. {Benz} and S.~{Krucker}, {\em \apj} {\bf 568}, 413 (March 2002),
  \href{http://arxiv.org/abs/astro-ph/0109027}{{\ttfamily astro-ph/0109027}}.

\bibitem{Benz-2003}
A.~O. {Benz} and P.~C. {Grigis}, {\em Advances in Space Research} {\bf 32},
  1035 (September 2003),
  \href{http://arxiv.org/abs/astro-ph/0308323}{{\ttfamily astro-ph/0308323}}.

\bibitem{Pauluhn:2006ut}
A.~{Pauluhn} and S.~K. {Solanki}, {\em \aap} {\bf 462}, 311 (January 2007),
  \href{http://arxiv.org/abs/astro-ph/0612585}{{\ttfamily astro-ph/0612585}}.

\bibitem{Bingert:2013}
{Bingert, S.} and {Peter, H.}, {\em A\&A} {\bf 550},   A30  (2013).

\bibitem{Klimchuk:2005nx}
J.~A. {Klimchuk}, {\em \solphys} {\bf 234}, 41 (March 2006),
  \href{http://arxiv.org/abs/astro-ph/0511841}{{\ttfamily astro-ph/0511841}}.

\bibitem{Klimchuk:2017}
J.~A. {Klimchuk}, {\em ArXiv e-prints}  (September 2017),
  \href{http://arxiv.org/abs/1709.07320}{{\ttfamily arXiv:1709.07320
  [astro-ph.SR]}}.

\bibitem{Bertolucci-2017}
S.~{Bertolucci}, K.~{Zioutas}, S.~{Hofmann} and M.~{Maroudas}, {\em Physics of
  the Dark Universe} {\bf 17}, 13 (September 2017),
  \href{http://arxiv.org/abs/1602.03666}{{\ttfamily arXiv:1602.03666
  [astro-ph.SR]}}.

\bibitem{Mondal-2020}
S.~Mondal, D.~Oberoi and A.~Mohan, {\em \apj} {\bf 895},   L39 (Jun 2020).

\bibitem{Thejappa-1991}
G.~Thejappa, {\em Solar Physics} {\bf 132}, 173  (1991).

\bibitem{Zhitnitsky:2020shd}
A.~Zhitnitsky, {\em J. Phys. G} {\bf 48},   065201  (2021),
  \href{http://arxiv.org/abs/2008.04325}{{\ttfamily arXiv:2008.04325
  [hep-ph]}}.

\bibitem{Liang:2021wjx}
X.~Liang and A.~Zhitnitsky (1 2021),
  \href{http://arxiv.org/abs/2101.01722}{{\ttfamily arXiv:2101.01722
  [hep-ph]}}.

\bibitem{Abbasi:2017rvx}
Telescope Array Project Collaboration, R.~Abbasi {\em et~al.}, {\em Phys. Lett.
  A} {\bf 381}, 2565  (2017).

\bibitem{Okuda_2019}
T.~Okuda, {\em Journal of Physics: Conference Series} {\bf 1181},   012067 (feb
  2019).

\bibitem{Gurevich_2001}
A.~V. Gurevich and K.~P. Zybin, {\em Physics-Uspekhi} {\bf 44}, 1119 (nov
  2001).

\bibitem{DWYER2014147}
J.~R. Dwyer and M.~A. Uman, {\em Phys. Rep.} {\bf 534}, 147   (2014), The
  Physics of Lightning.

\bibitem{Fischer:2018niu}
H.~Fischer, X.~Liang, Y.~Semertzidis, A.~Zhitnitsky and K.~Zioutas, {\em Phys.
  Rev. D} {\bf 98},   043013  (2018),
  \href{http://arxiv.org/abs/1805.05184}{{\ttfamily arXiv:1805.05184
  [hep-ph]}}.

\bibitem{Liang:2018ecs}
X.~Liang and A.~Zhitnitsky, {\em Phys. Rev. D} {\bf 99},   023015  (2019),
  \href{http://arxiv.org/abs/1810.00673}{{\ttfamily arXiv:1810.00673
  [hep-ph]}}.

\bibitem{Liang:2019lya}
X.~Liang, A.~Mead, M.~S.~R. Siddiqui, L.~Van~Waerbeke and A.~Zhitnitsky, {\em
  Phys. Rev. D} {\bf 101},   043512  (2020),
  \href{http://arxiv.org/abs/1908.04675}{{\ttfamily arXiv:1908.04675
  [astro-ph.CO]}}.

\bibitem{Budker:2019zka}
D.~Budker, V.~V. Flambaum, X.~Liang and A.~Zhitnitsky, {\em Phys. Rev. D} {\bf
  101},   043012  (2020), \href{http://arxiv.org/abs/1909.09475}{{\ttfamily
  arXiv:1909.09475 [hep-ph]}}.

\bibitem{Liang:2020mnz}
X.~Liang, E.~Peshkov, L.~Van~Waerbeke and A.~Zhitnitsky, {\em Phys. Rev. D}
  {\bf 103},   096001  (2021),
  \href{http://arxiv.org/abs/2012.00765}{{\ttfamily arXiv:2012.00765
  [hep-ph]}}.

\bibitem{Freese:1987wu}
K.~Freese, J.~A. Frieman and A.~Gould, {\em Phys. Rev.} {\bf D37}, 3388
  (1988).

\bibitem{Freese:2012xd}
K.~Freese, M.~Lisanti and C.~Savage, {\em Rev. Mod. Phys.} {\bf 85}, 1561
  (2013), \href{http://arxiv.org/abs/1209.3339}{{\ttfamily arXiv:1209.3339
  [astro-ph.CO]}}.

\bibitem{Oaknin:2004mn}
D.~H. {Oaknin} and A.~R. {Zhitnitsky}, {\em Phys. Rev. Lett.} {\bf 94},
  101301 (March 2005), \href{http://arxiv.org/abs/hep-ph/0406146}{{\ttfamily
  hep-ph/0406146}}.

\bibitem{Zhitnitsky:2006tu}
A.~{Zhitnitsky}, {\em \prd} {\bf 76},   103518 (November 2007),
  \href{http://arxiv.org/abs/astro-ph/0607361}{{\ttfamily astro-ph/0607361}}.

\bibitem{Forbes:2006ba}
M.~{McNeil Forbes} and A.~R. {Zhitnitsky}, {\em \jcap} {\bf 1},   023 (January
  2008), \href{http://arxiv.org/abs/astro-ph/0611506}{{\ttfamily
  astro-ph/0611506}}.

\bibitem{Lawson:2007kp}
K.~{Lawson} and A.~R. {Zhitnitsky}, {\em \jcap} {\bf 1},   022 (January 2008),
  \href{http://arxiv.org/abs/0704.3064}{{\ttfamily arXiv:0704.3064}}.

\bibitem{Forbes:2008uf}
M.~M. {Forbes} and A.~R. {Zhitnitsky}, {\em \prd} {\bf 78},   083505 (October
  2008), \href{http://arxiv.org/abs/0802.3830}{{\ttfamily arXiv:0802.3830}}.

\bibitem{Forbes:2009wg}
M.~M. {Forbes}, K.~{Lawson} and A.~R. {Zhitnitsky}, {\em \prd} {\bf 82},
  083510 (October 2010), \href{http://arxiv.org/abs/0910.4541}{{\ttfamily
  arXiv:0910.4541}}.

\bibitem{Fraser:2014wja}
G.~W. Fraser, A.~M. Read, S.~Sembay, J.~A. Carter and E.~Schyns, {\em Mon. Not.
  Roy. Astron. Soc.} {\bf 445}, 2146  (2014),
  \href{http://arxiv.org/abs/1403.2436}{{\ttfamily arXiv:1403.2436
  [astro-ph.HE]}}.

\bibitem{Ge:2020cho}
S.~Ge, H.~Rachmat, M.~S.~R. Siddiqui, L.~Van~Waerbeke and A.~Zhitnitsky (4
  2020), \href{http://arxiv.org/abs/2004.00632}{{\ttfamily arXiv:2004.00632
  [astro-ph.HE]}}.

\bibitem{Budker:2020mqk}
D.~Budker, V.~V. Flambaum and A.~Zhitnitsky (3 2020),
  \href{http://arxiv.org/abs/2003.07363}{{\ttfamily arXiv:2003.07363
  [hep-ph]}}.

\bibitem{Figueroa:2021bab}
N.~L. Figueroa, D.~Budker and E.~M. Rasel, {\em Quantum Sci. Technol.} {\bf 6},
    034004  (2021), \href{http://arxiv.org/abs/2103.08715}{{\ttfamily
  arXiv:2103.08715 [astro-ph.CO]}}.

\bibitem{Gorham:2016zah}
ANITA Collaboration, P.~W. Gorham {\em et~al.}, {\em Phys. Rev. Lett.} {\bf
  117},   071101  (2016), \href{http://arxiv.org/abs/1603.05218}{{\ttfamily
  arXiv:1603.05218 [astro-ph.HE]}}.

\bibitem{Gorham:2018ydl}
ANITA Collaboration, P.~W. Gorham {\em et~al.}, {\em Phys. Rev. Lett.} {\bf
  121},   161102  (2018), \href{http://arxiv.org/abs/1803.05088}{{\ttfamily
  arXiv:1803.05088 [astro-ph.HE]}}.

\bibitem{Liang:2021rnv}
X.~Liang and A.~Zhitnitsky (5 2021),
  \href{http://arxiv.org/abs/2105.01668}{{\ttfamily arXiv:2105.01668
  [hep-ph]}}.

\end{thebibliography}

\end{document}